\documentclass{article} 
\usepackage[a4paper, margin=1.5in]{geometry}
\usepackage{amssymb}
\usepackage{amsmath}
\usepackage{graphicx}
\usepackage{subfigure}
\usepackage{makeidx}
\usepackage{multicol}
\usepackage{algorithm}
\usepackage{booktabs}
\usepackage{changepage}
\usepackage{bm}

\begin{document}

\title{Sparse Bayesian State-Space and Time-Varying Parameter Models\footnote{Also appears as a chapter in the Handbook of Bayesian Variable Selection \cite{tad-van:handbook}.}}
\author{Sylvia Frühwirth-Schnatter$^1$ \and Peter Knaus$^1$}
\date{%
    $^1$Vienna University of Economics and Business%
}

\maketitle

%
\newcommand{\alphav}{\boldsymbol{\alpha}} 
\newcommand{\Am}{{\mathbf A}}
\newcommand{\Amul}{\boldsymbol{A}} 
\newcommand{\ataupr}{b^{\tau}}  
\newcommand{\axipr}{b^{\xi}}    

\newcommand{\betac}{\beta} 
\newcommand{\betaci}[1]{\betac_{#1}}
\newcommand{\betad}{p}    
\newcommand{\Betapr}[1]{\mathcal{BP}\left(#1\right)}
\newcommand{\betar}{\boldsymbol{\betac}} 
\newcommand{\betav}{\betar} 
\newcommand{\betavtilde}{\tilde \betav}
\newcommand{\betavred}{\betav_{\deltac,\gamma}}
 \newcommand{\btildev}[1]{\tilde{\betav}_{#1}}
\newcommand{\Betafun}[1]{B(#1)} 
\newcommand{\Betadis}[1]{\mathcal{B}eta\left(#1\right)}
\newcommand{\Bino}[1]{\mathcal{B}in\left(#1\right)}
\newcommand{\bfz}{{\mathbf{0}}}
\newcommand{\bfzmat}{{\mathbf{O}}}
\newcommand{\Bm}{{\mathbf B}}
\newcommand{\Bv}[1]{\mathbf{B}_ #1}
\newcommand{\cm}{\mathbf{c}}
\newcommand{\Cov}[1]{Cov (#1)}
\newcommand{\deltac}{\delta} 
\newcommand{\deltav}{\boldsymbol{\deltac}}
\newcommand{\Det}[1]{|#1|}  
\newcommand{\Diag}[1]{\mbox{\rm Diag}\left(#1\right)}  
\newcommand{\dimmat}[2]{(#1\times #2)} 
\newcommand{\dimy}{\ymd}
\newcommand{\dirac}[1]{\delta_ #1}
\newcommand{\Dm}{{\mathbf D}}
\newcommand{\Dmul}{\boldsymbol{D}} 

\newcommand{\e}{\mbox{\rm e}}
\newcommand{\Ew}[1]{\mbox{\rm E}[#1]}   
\newcommand{\error}{\varepsilon} 
\newcommand{\errorv}{\boldsymbol{\error}}
\newcommand{\etav}{{\boldsymbol{\eta}}} 
\newcommand{\Exp}[1]{\mathcal{E}xp\left(#1\right)}
\newcommand{\Fd}[1]{\mbox{\rm F}\left(#1\right)}
\newcommand{\gammac}{\gamma} 
\newcommand{\gammav}{\boldsymbol{\gammac}} 
\newcommand{\Gamfun}[1]{\Gamma (#1)} 
\newcommand{\Gammad}[1]{ \mathcal{G}\left(#1\right)}
\newcommand{\Gammainv}[1]{\mathcal{IG}  \left(#1\right)}
\newcommand{\GIG}[3]{ \mathcal{GIG}\left(#1,#2,#3\right)}
\newcommand{\im}[1]{^{(#1)}}
\newcommand{\indicset}[2]{\mathbb{I}_{#1}( #2)} 
\newcommand{\indicevent}[1]{\mathbb{I}\{#1\}} 
\newcommand{\identm}{I}  
\newcommand{\identy}[1]{{\identm}_{#1}} 
\newcommand{\indic}[1]{\mathbb{I} (#1)}
\newcommand{\kappashr}{\rho}
\newcommand{\kfQc}{\theta}  
\newcommand{\kfQ}{{\mathbf{Q}}}  
\newcommand{\kfxc}{\betac}  
\newcommand{\kfx}{\betav}  
\newcommand{\kfzm}{\mathbf{z}}  
\newcommand{\kfwc}{w}
\newcommand{\kfw}{{\mathbf{\kfwc}}}  
\newcommand{\lambdav}{\boldsymbol{\lambda}}
\newcommand{\LPS}{{\mbox{\rm LPDS}}}
\newcommand{\LPSo}[1]{\LPS^{\star}_{#1}}
\newcommand{\mean}[1]{\overline{#1}}
\newcommand{\MVAR}{\ymd}
\newcommand{\new}{^{\rm \tiny new}}
\newcommand{\newarg}[1]{^{(#1, new)}}
\newcommand{\Normal}[1]{ \mathcal{N}\left(#1\right)}
\newcommand{\Normult}[2]{ \mathcal{N} _{#1}\left(#2\right)}
\newcommand{\old}{^{old}}
\newcommand{\oldarg}[1]{^{(#1, old)}}
\newcommand{\omegav}{\boldsymbol{\omega}}
\newcommand{\ones}[1]{\mathbf{1}_{#1}}   
\newcommand{\pdim}[1]{p_{#1}}
\newcommand{\Phim}{\boldsymbol{\Phi}}
\newcommand{\phitau}{\phi^{\tau}}
\newcommand{\phixi}{\phi^{\xi}}
\newcommand{\pimix}[1]{\pi_{#1}}
\newcommand{\Pm}{\mathbf{P}}
\newcommand{\Probsym}{\mbox{\rm P}}
\newcommand{\Prob}[1]{\Probsym (#1)}
\newcommand{\pVAR}{r}
\newcommand{\Scov}[1]{{\mathbf S}_{#1}} 
\newcommand{\sigmaerr}{\sigma^2}  
\newcommand{\Sigmam}{\boldsymbol{\Sigma}}  
\newcommand{\spike}{\mbox{\tiny spike}}
\newcommand{\slab}{\mbox{\tiny slab}}
\newcommand{\Student}[2]{t _{#1} \left(#2\right)}
\newcommand{\Studentnu}[1]{t _{#1}}
\newcommand{\sv}[1]{s^2_{#1}} 
\newcommand{\tauv}{\boldsymbol{\tau}} 
\newcommand{\thc}{\vartheta} 
\newcommand{\thmod}{{\mathbf{\boldsymbol{\thc}}}} 
\newcommand{\thetav}{{\mathbf{\boldsymbol{\theta}}}} 
\newcommand{\TPB}[1]{\mathcal{TPB}\left(#1\right)}
\newcommand{\trans}[1]{#1 ^{\top}} 
\newcommand{\Uhyp}[1]{U \left(#1\right)}
\newcommand{\Uniform}[1]{\mathcal{U}\left[#1\right]}
\newcommand{\Var}[1]{Var \left(#1\right) }
\newcommand{\verror}{\sigma^2}
\newcommand{\Xbeta}{{\mathbf x}} 
\newcommand{\Xbetamat}{{\mathbf X}}  
\newcommand{\xiFsqr}{\psi}
\newcommand{\xiF}{\xiFsqr^2}
\newcommand{\xiv}{\boldsymbol{\xi}} 
\newcommand{\xm}{\Xbeta}
\newcommand{\ym}{{\mathbf y}} 
\newcommand{\ymd}{q} 
\newcommand{\tr}{{\tiny \mbox{\rm tr}}}
\newcommand{\ytr}{\ym^{\tr}}
\newcommand{\aphi}{a_\phi} 
\newcommand{\bphi}{b_\phi} 
\newcommand{\Bsv}{B_\sigma} 


\newcommand{\remove}[1]{} 
\newcommand{\Pc}[1]{#1} 
\newcommand{\Pcm}[1]{{#1}} 
\newcommand{\PcS}[1]{#1} 
\newcommand{\comment}[1]{#1} 
\newcommand{\commentS}[1]{#1} 
\newcommand{\typo}[1]{#1} 

\begin{abstract}
  In this chapter, we review variance selection for time-varying parameter (TVP) models for univariate and multivariate time series within a Bayesian framework.
We show how both continuous as well as discrete spike-and-slab shrinkage priors can be transferred from variable selection for regression models to variance selection for TVP models by using a non-centered parametrization. We discuss efficient MCMC estimation and provide an application to US inflation modeling.
\end{abstract}

\section{Introduction}   

Time-varying parameter (TVP) \comment{models and,  more generally}, state space models are widely used in time series analysis to
deal with model coefficients that  change over time.
 \Pc{This ability to capture gradual changes is one of state space models greatest advantages. The flipside of this high degree of flexibility\Pcm{, however,} is that they run} the risk of overfitting with a growing number of coefficients, as many of them might, in reality, be constant over the entire observation period. This will be exemplified in the present chapter with an economic application. We will model US inflation through a TVP Phillips curve, where, out of 18 potentially time-varying coefficients, only a single one actually  changes over time. We will show that allowing static coefficients to  be time-varying leads to a considerable loss of statistical efficiency, both in uncertainty quantification for the parameters and forecasting future \comment{time series} observations. We will also  show that \Pc{substantial} statistical efficiency can be gained by applying a Bayesian estimation strategy that is able to single out  parameters that are indeed constant or even insignificant.

 Identifying \comment{constant} coefficients in a TVP model   amounts to a {\em variance selection} problem, involving a decision \comment{on} whether the variances of the shocks driving the dynamics of  a  time-varying  parameter  are equal to zero.  Variance selection in latent variable models is known to be a non-regular problem
within the framework of classical statistical hypothesis testing \cite{har:for}. The introduction of  shrinkage priors
 for  the variances of a TVP model within a Bayesian  framework has proven to be a very useful strategy which is capable of  automatically reducing time-varying coefficients to static ones if the model \comment{overfits.}

In pioneering work, \cite{fru-wag:sto} reformulated the {\em variance selection} problem for state space models as a
 {\em variable selection} problem in the so-called non-centered parametrization of the TVP model.  This insight established
a general strategy  for  extending  shrinkage priors from  standard  regression analysis  to this more general framework. For variance selection in \lq\lq sparse\rq\rq\ state space and TVP models, \cite{fru-wag:sto} employed discrete spike-and-slab priors, \cite{bel-etal:hie} relied on the  Bayesian Lasso prior,
    \cite{bit-fru:ach} applied the normal-gamma prior of \cite{gri-bro:inf} and \cite{cad-etal:tri} introduced the triple gamma prior, which is related to the normal-gamma-gamma prior
   \cite{gri-bro:hie} and contains the horseshoe prior \cite{car-etal:hor} as a special case.

    The present chapter  reviews this literature, starting in Section~\ref{ssm:sec:intro} with univariate time-varying parameter models. \Pc{In particular,} we will demonstrate that the commonly used inverse gamma prior on the process variances prevents variance selection. Using a ridge prior in the non-centered TVP model
     instead of \Pc{an} inverse gamma \Pc{prior} provides a simple, yet useful alternative. The ridge prior can be translated into a gamma prior for the variances and leads to more reliable uncertainty quantification in parameter estimation and forecasting for sparse state space models.
    Starting from the ridge prior, continuous shrinkage priors for variance selection are discussed in Section~\ref{ssm:sec:shrink}, whereas
    Section~\ref{ssm:sec:BVS} discusses discrete spike-and-slab priors. In both sections, we also review strategies for efficient Markov chain Monte Carlo (MCMC) estimation\Pc{,} which is even more challenging for state space model\Pc{s} than for standard regression models.
   Section~\ref{ssm:sec:mult} discusses extensions to multivariate time series, including TVP Bayesian vector autoregressive  models and TVP Cholesky stochastic volatility models,
    \comment{shows} how to compare various shrinkage priors through log predictive density scores and addresses the issues of classifying coefficients into dynamic or \comment{constant} ones. Section~\ref{section:ssm:conclude} concludes with a \PcS{brief} discussion.

\section{Univariate time-varying parameter models \label{ssm:sec:intro}}

\subsection{Motivation and model definition \label{ssm:sec:motivate}}

In this section, we consider time-varying parameter (TVP) models for a univariate time series $y_t$. For $t = 1, \ldots, T$, we have that
\begin{equation} \label{ssm:eq:centeredpar}
\begin{aligned}
& \kfx_{t}  = \kfx_{t-1} + \kfw_{t}, \qquad   \kfw_t  \sim \Normult{\betad}{\bfz, \kfQ},\\
& y_{t}=   \Xbeta_t \kfx_{t}  +  \error_{t} , \qquad \error_{t} \sim \Normal{0,\sigma^2},  
\end{aligned}
\end{equation}
where  \comment{$\kfx_{t}=\trans{(\beta_{1t}, \ldots, \beta_{pt})}$ is a latent state variable
and the covariance $\kfQ=\Diag{\theta_1, \ldots, \theta_\betad}$ of the innovations $\kfw_{t}$ is diagonal.}
$ \Xbeta_t = (\typo{x_{1 t}},  \ldots, \typo{x_{\betad t}})$
is a $\betad$-dimensional row vector containing the explanatory variables at time $t$. The variables $\typo{x_{j t}}$ can be exogenous {(i.e. determined outside the model)} control variables and/or be equal to lagged values of $y_{t}$.
Usually, one of the variables, say $\typo{x_{1 t}}$, corresponds to the intercept, but an intercept need not be present.
\comment{In Section~\ref{ssm:sec:mult}}, this approach is extended to multivariate time series $\ym_t$.

\Pc{To fully specify the model}, a distribution has to be \Pc{defined}
for the initial value $\kfx_{0}$ of the state process, with a typical choice being a normal distribution, e.g. $\kfx_{0} \sim \Normult{\betad}{\betav, \kfQ}$, with initial expectation $\betav = (\beta_1, \ldots, \beta_\betad)^\top$.
An alternative choice is to assume a diffuse prior with  fixed initial expectation and a very uninformative prior covariance matrix, e.g. $\kfx_{0} \sim \Normult{\betad}{\bfz,  10^5 \cdot\identy{\betad} }$ where $\identy{\betad}$ is the $\betad$-dimensional identity matrix. However, such a choice is not recommended for TVP models where overfitting presents a concern.

\begin{figure}[t!]
\centering
\includegraphics[width=0.8\linewidth]{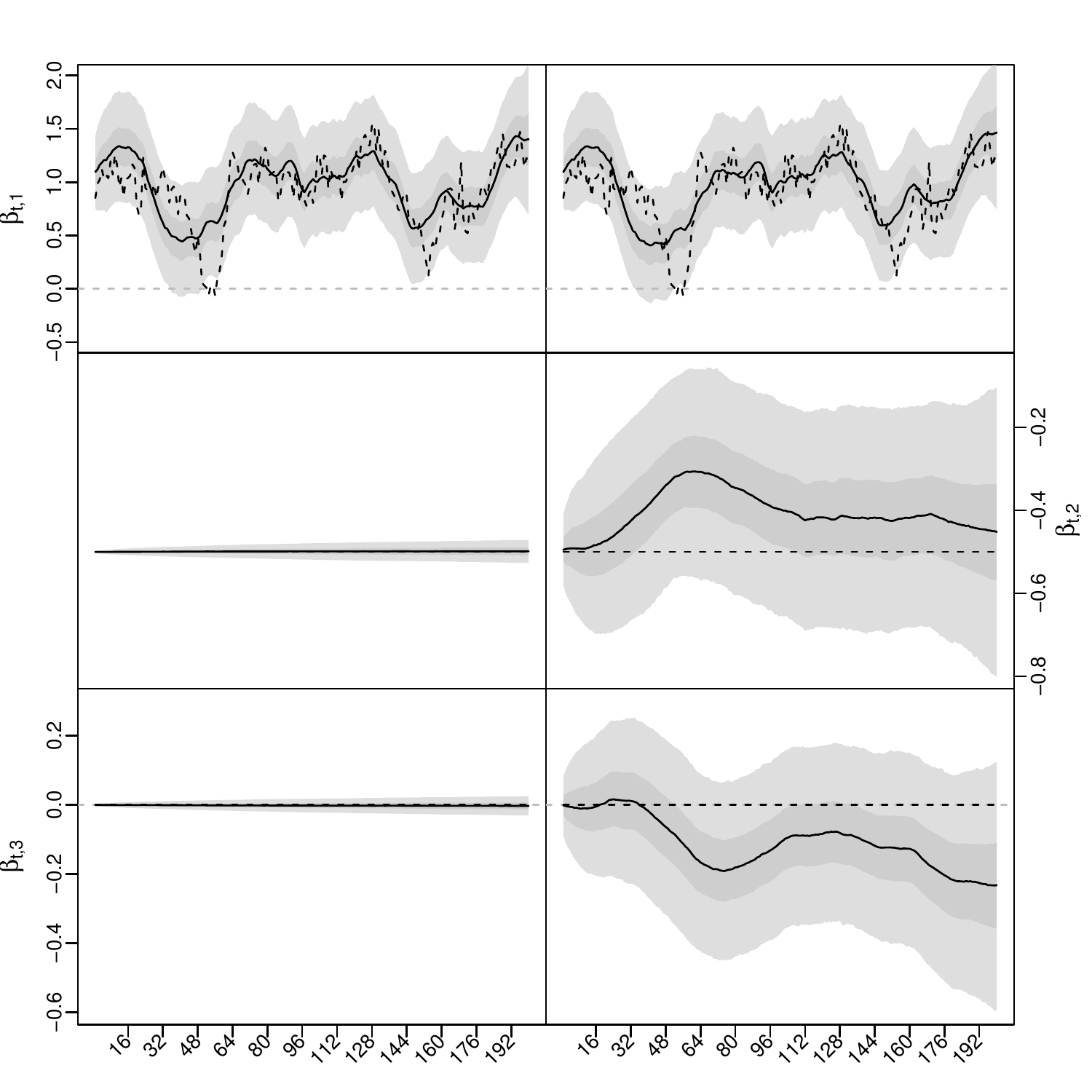} 
\caption{\comment{Recovery of} the hidden parameters of a TVP model for simulated data using the Kalman filter and smoother
 in an overfitting TVP model with $\theta_2=\theta_3=10^{-6}$  (left-hand side) and   $\theta_2=\theta_3=0.001$ (right-hand side). \Pcm{The gray shaded regions represent pointwise 95\% and 50\% credible intervals, respectively, while the black solid line represents the pointwise median. The black dashed line represents the true hidden parameter values.}}
\label{ssm:fig1}
\end{figure}

The goal is to recover the unobserved state process  $\kfx_{0}, \ldots, \kfx_{T}$ given the observed time series
 $\ym=(y_1, \ldots,y_T)$.  If  $\betav$, $\kfQ$ and $\sigma^2$ were known, this is easily achieved by the famous Kalman filter and smoother \cite{kal:new}. 
 For illustration,  a time series $y_t$ is generated from model (\ref{ssm:eq:centeredpar}) with $T=200$, $\betad=3$,
 $\typo{x_{1 t}}= 1$, $\typo{x_{j t}} \sim  \Normal{0,1}$, $j=2,3$, $\sigma^2=1$, $(\beta_1, \beta_2,\beta_3)=(1,-0.5,0)$
 and  $(\theta_1, \theta_2,\theta_3)=(0.02,0,0)$.  The paths of the hidden process $ \kfx_{t}$ are reconstructed  using the Kalman filter and smoother based on the true values of $\betav$, $\theta_1$ and $\sigma^2$ and very small values for
 $\theta_2=\theta_3=10^{-6}$
 and compared to the true paths \Pc{in the left-hand side of Figure~\ref{ssm:fig1}}. Since the marginal posterior of $\kfx_{t}|\ym$ is a Gaussian distribution for each $t$, point-wise credible regions  for  $\kfx_{t}$ are easily obtained which are very helpful for uncertainty quantification. Although the TVP model used for estimation \comment{overfits}, 
 the Kalman smoother is rather accurate in recovering the true paths and clearly indicates that the last coefficients are constant,
 {\em assuming} that $\theta_2$ and $\theta_3$ are very close to 0.

 However, in real-world applications, the  variances $\theta_j$ are unknown and \comment{estimated} from the observed time series, together with the entire path $\kfzm=(\kfx_{0}, \ldots, \kfx_{T})$.
As  
\comment{evident from} the Kalman filter, the variances $\kfQ$ of the innovations $\kfw_t $ play an important role in quantifying the loss from propagating  the  filtering density
$  \kfx_{t-1}| \comment{\ym^{t-1}} \sim 
\Normult{\betad}{{\mathbf m}_{t-1|t-1}, \Pm_{t-1|t-1}},$
\comment{given $\ym^{t-1}=(y_1, \ldots, y_{t-1})$},
into the future to forecast \comment{$\kfx_{t}$}: $$  \kfx_{t}|\comment{\ym^{t-1}} \sim \Normult{\betad}{{\mathbf m}_{t-1|t-1}, \Pm_{t-1|t-1}+\kfQ}.$$
 A comparably minor change of $\kfQ$ can have a strong effect on uncertainty quantification. For instance,  assuming  $\theta_2=\theta_3=0.001$ (instead of \comment{$10^{-6}$})
\PcS{for the simulated data} has a huge effect on the recovered paths, as shown in \comment{the right-hand side of} Figure~\ref{ssm:fig1}. Not only are the \comment{credible intervals}  much \comment{broader}, we can also no longer be sure if the two coefficients $\beta_{2t}$ and $\beta_{3t}$ are time-varying \comment{or constant}.


In a maximum likelihood 
framework, the Kalman filter is used to compute the likelihood function\Pc{,} which is maximized to obtain estimates of \comment{$\theta_1, \ldots, \theta_\betad$, 
$\sigma^2,$ and $\beta_1, \ldots, \beta_\betad$ (if the initial means  are 
 unknown)}. \comment{Reconstructing $\kfzm=(\kfx_{0}, \ldots, \kfx_{T})$} then operates conditional on these estimates, see e.g. \cite{har:for}.


For Bayesian inference, priors are chosen for 
$\theta_1, \ldots, \theta_\betad$, \comment{$\sigma^2$, and
$\beta_1, \ldots, \beta_\betad$}.
 Given  time series \comment{observations $\ym=(y_1, \ldots, y_T)$}, the joint posterior distribution $p(\kfzm,\betav, \kfQ,\sigma^2|\ym)$ is the object of interest from which marginal posteriors $p(\kfx_{t}|\ym)$ are derived for each $t$. These can be used for uncertainty quantification as in Figure~\ref{ssm:fig1}, while also taking uncertainty in the model parameters into account.
Different algorithms have been developed to sample from the joint posterior $p(\kfzm,\betav, \kfQ,\sigma^2|\ym)$, in particular two-block Gibbs samplers that alternate between drawing from $p(\kfzm|\betav, \kfQ,\sigma^2,\ym)$ using forward-filtering, backward-sampling (FFBS) \cite{car-koh:ong,fru:dat} and  drawing from $p(\betav, \kfQ,\sigma^2| \kfzm,\ym)$.

Both maximum likelihood (ML) 
and Bayesian inference work well for TVP models where all state variables $\kfxc_{jt}$ are dynamic.\remove{in the sense that temporal changes of $\kfxc_{jt}$ contribute to explaining the variation of $y_t$.  In such a case the dynamic coefficient of determination
\begin{eqnarray*}
R_j=\frac{\Ew{x^2_{t j}} \theta_j}{\sum_{l=1}^\betad \Ew{x^2_{t l}} \theta_l + \sigma^2}
\end{eqnarray*}
is not negligible for all coefficients $\kfxc_{jt}$.} If one of the variances $\theta_j$ is equal to 0, ML estimation leads to a non-regular testing problem, since the true value lies on the boundary of the parameter space \cite{har:for}.
\comment{As opposed to this,} Bayesian inference is able to deal with such sparse TVP models and, more generally, sparse state space models. The two main challenges from the Bayesian perspective are \PcS{the choice of an appropriate prior for the variances $\theta_j$} %
and computational challenges with regards to \PcS{efficient} MCMC estimation.

\subsection{The inverse gamma versus the ridge prior\label{ssm:sec:noncen}}

A popular prior choice  for the process variance $\theta_j$ is
 the inverse gamma  distribution,
\begin{eqnarray} \label{gaminv}
 \theta_j  \sim  \Gammainv{s_0,S_0},
 \end{eqnarray}
 which is often applied with very small hyperparameters, e.g. $s_0=S_0=0.001$ \cite{pet-etal:dyn}.
Given the latent process $(\beta_{j0}, \ldots, \beta_{jT})$, this  prior is conditionally conjugate in the \comment{so-called} centered parametrization (\ref{ssm:eq:centeredpar}),
 since the \comment{density} 
 $p(\beta_{j0}, \ldots, \beta_{jT}|\theta_j)$ is the kernel of an inverse gamma  distribution.
 Hence,  prior (\ref{gaminv}) leads to  an  inverse gamma  posterior distribution  $p(\theta_j| \beta_{j0}, \ldots, \beta_{jT})$.
 However,  this prior performs poorly when dealing with a sparse TVP model,
 it is bounded away from zero, making it incapable of inducing strong shrinkage \cite{fru-wag:sto}.

The effect of choosing a specific  prior becomes more apparent  when we rewrite
model (\ref{ssm:eq:centeredpar}) in  the non-centered parametrization introduced in \cite{fru-wag:sto}:
\begin{equation}  \label{ssm:eq:noncenteredpar}
\begin{aligned}
&\btildev{t} =\btildev{t-1} + \tilde{\kfw}_{t}, \qquad \tilde{\kfw}_{t} \sim  \Normult{\betad}{\bfz, \identy{\betad}}, \\
&y_t= \Xbeta_t   \betav +  \Xbeta_t   \Diag{\sqrt \theta_1 , \ldots, \sqrt \theta_\betad} \btildev{t}
+  \error_t, \quad  \error_t \sim \Normal{0,\sigma^2},
\end{aligned}
\end{equation}
with initial distribution $ \btildev{0} \sim \Normult{\betad}{\bfz, \identy{\betad}} $.   A  linear transformation connects the two parametrizations:
\begin{eqnarray}\label{ssm:eq:solve}
\beta_{jt} = \beta_j + \sqrt \theta_j  \tilde \beta_{jt}, \quad  t=0,\ldots,T, \quad j=1, \ldots, \betad.
\end{eqnarray}
\comment{Evidently,} both representations are equivalent, and
  we can specify a prior either  on
the variances $\theta_j$ in (\ref{ssm:eq:centeredpar}) or \comment{on} the scale parameters $\sqrt \theta_j$ in (\ref{ssm:eq:noncenteredpar}).
 \comment{Since the conjugate prior for $\sqrt \theta _j$ in the
non-centered parametrization (\ref{ssm:eq:noncenteredpar}) is the normal distribution,
 the scale parameter $ \sqrt \theta _j$ is  assumed to be Gaussian:
 \begin{eqnarray} \label{ssm:normal:sigma}
  \sqrt \theta _j | \sigma^2  \sim \Normal{0,\sigma^2 \comment{\tau}} \quad \Leftrightarrow  \quad \theta_j| \sigma^2  \sim  \Gammad{\frac{1}{2},\frac{1}{2  \comment{\tau} \sigma^2}}.
 \end{eqnarray}
 Here,  $\sqrt \theta _ j \in \mathbb{R}$ 
 is allowed  to  take on both positive and negative values. This implies that $\theta_j =(\sqrt \theta_ j)^2$ follows a re-scaled $\chi^2_1$-distribution.} \cite{fru:com_eff} introduced such a shrinkage prior (with fixed scale parameter $\tau$) for the process variance in a univariate TVP model (that is  $\betad=1$),  and  \cite{fru-wag:sto} extended  this idea to state space models with $\betad>1$.
  Alternatively, it \comment{can be assumed} that the prior scale is independent of $\sigma^2$, i.e.
  \begin{eqnarray} \label{ssm:normal}
  \sqrt \theta _j   \sim \Normal{0, \comment{\tau}} \quad \Leftrightarrow  \quad \theta_j  \sim  \Gammad{\frac{1}{2},\frac{1}{2 \comment{\tau}}}.
 \end{eqnarray}
 \comment{As shown by \cite{mor-etal:var}, such a prior has certain advantages compared to (\ref{ssm:normal:sigma}) and allows} \Pc{for the introduction of} stochastic volatility in model (\ref{ssm:eq:centeredpar}), \comment{see \cite{kna-etal:shr_tim} and} Section~\ref{ssm:sec:sv}.

 \begin{figure}[t!]
  \centering
  \includegraphics[width=\linewidth]{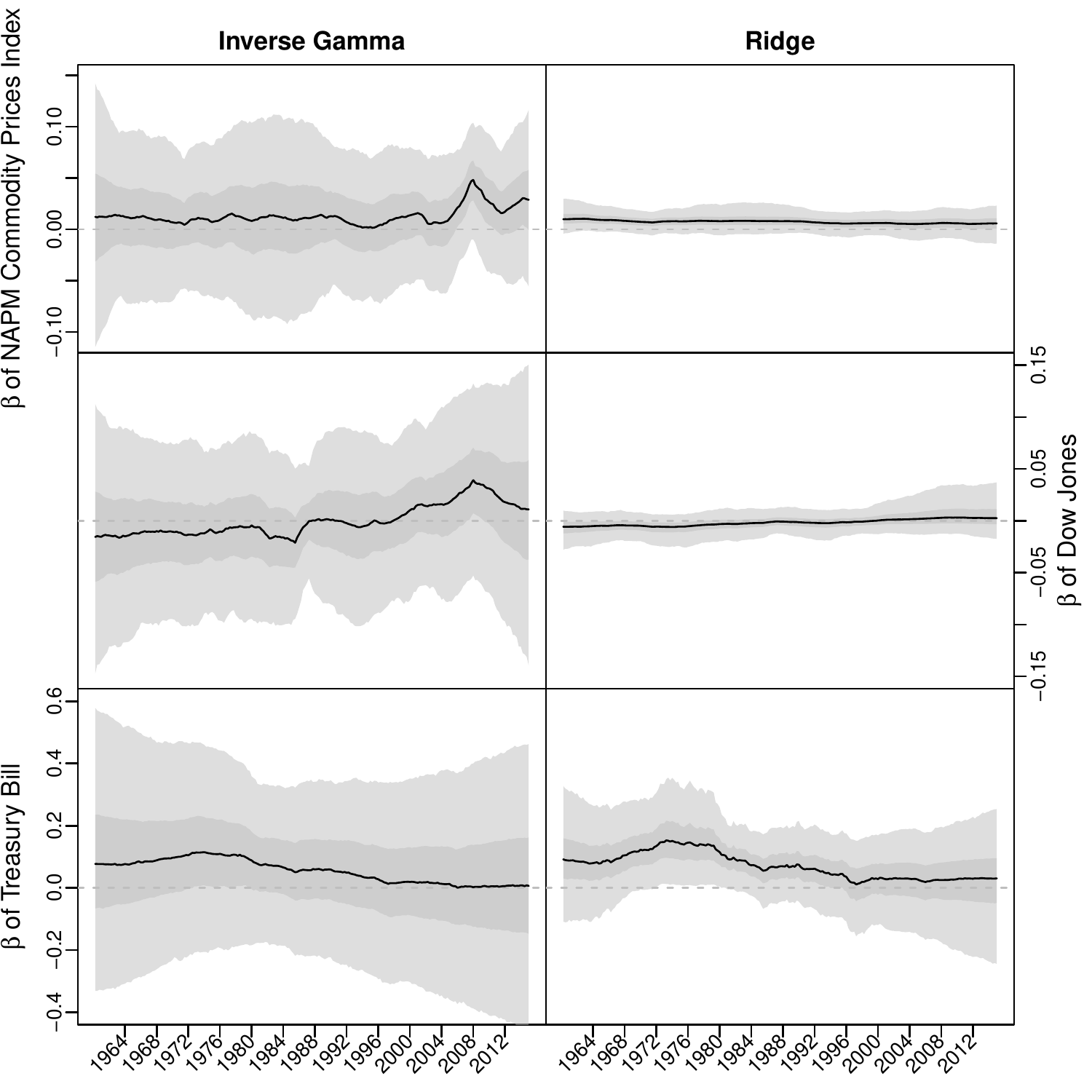}
  \caption{\comment{Recovery of} the time-varying parameters for the inflation data
   under the inverse gamma \comment{$\theta_j  \sim  \Gammainv{0.001,0.001}$} and
   under the ridge prior \comment{$\theta_j  \sim  \Gammad{0.5,10}$}.
   The gray \comment{shaded regions} represent pointwise 95\% and 50\% credible intervals, respectively, while the black line represents the pointwise median.}
  \label{ssm:fig2}
  \end{figure}

From the viewpoint of variable selection, prior (\ref{ssm:normal}) is a  ridge prior in a standard regression model, conditional on the hidden path
\comment{$\kfzm=(\tilde \kfx_{0}, \ldots, \tilde \kfx_{T})$.} 
Many variable selection priors have been introduced for \Pc{standard regression models} (albeit with known rather than latent regressors), \comment{see \cite{bha-etal:las} for a recent review.}
Given the non-centered parametrization (\ref{ssm:eq:noncenteredpar}),  any of these priors can be, in principle, applied in the context of sparse TVP and state space models for variance selection. And, indeed, the literature has seen an increasing number of papers \comment{following this lead}
 \cite{bel-etal:hie,bit-fru:ach,cad-etal:tri,fru:com_eff,fru-wag:sto}.


Shrinking $\theta_{j}$ toward the boundary value  is achieved  by shrinking  $  \sqrt \theta _j $   toward 0 (which is an interior point of the parameter space in the non-centered parametrization). For a sparse state space model,  prior (\ref{ssm:normal}) substitutes the inverse gamma prior (\ref{gaminv}) with a gamma prior.
 This change in the prior specification is negligible for truly dynamic models, where the posterior distribution
$p(\kfzm,\betav, \kfQ,\sigma^2|\ym)$ is fairly robust to prior choices $p(\theta_j)$,
but has a considerable effect on uncertainty quantification for the unknown path $\kfzm$ for a sparse state space model.
\comment{This} is illustrated in Figure~\ref{ssm:fig2}, where the gamma prior
\comment{$\theta_j  \sim  \Gammad{0.5,10}$} 
is compared to the  inverse gamma 
\comment{prior $\theta_j  \sim  \Gammainv{0.001,0.001}$}
for the inflation data that will be discussed in detail in  Section~\ref{ssm:sec:inf:shrink}.

  \begin{algorithm}[t!]
\begin{enumerate}
  \item[(a)] sample  the latent variables
 $\kfzm=(\tilde \kfx_{0}, \ldots, \tilde \kfx_{T})$ conditional on the  model parameters
 $\alphav=( \beta_1, \ldots, \beta_\betad , \sqrt\theta_1, \ldots, \sqrt\theta_\betad )$ and
 $\sigma^2$ from \comment{$\kfzm| \alphav, \sigma^2, \ym $}, using e.g. FFBS;
\item[(b)]\comment{sample $(\alphav, \sigma^2)$ conditional on $\kfzm$}:
\begin{enumerate}
\item[(b-1)]  sample $\sigma^2 $, respectively,  from the inverse gamma density
   $\sigma^2 | \kfzm, \ym $  or   $\sigma^2 | \alphav, \kfzm, \ym $   depending on whether the ridge priors' scale
   depends on $\sigma^2$ or not;
   \item[(b-2)] \comment{sample $\alphav$ 
       from the multivariate Gaussian $\alphav| \sigma^2 , \kfzm, \ym $.}
\end{enumerate}
\end{enumerate}
 \caption{MCMC sampling in the non-centered parametrization of a TVP model under the ridge prior.}\label{ssm:algo1}
\end{algorithm}

\subsection{Gibbs sampling in the non-centered parametrization}

A two-block Gibbs sampler is available to sample   the latent variables
 $\kfzm=(\tilde \kfx_{0}, \ldots, \tilde \kfx_{T})$ and the  model parameters $\alphav=(\beta_1, \ldots, \beta_\betad , \theta_1, \ldots, \theta_\betad)$ and $\sigma^2$ in the non-centered parametrization, see Algorithm~\ref{ssm:algo1}.
 In step~(b), if the  prior scale \comment{in (\ref{ssm:normal:sigma}) depends
 on $\sigma^2$ and, similarly, $\beta_j | \sigma^2  \sim \Normal{0,\sigma^2 \comment{\tau}}$}, then, conditional on $\kfzm$, the non-centered parametrization (\ref{ssm:eq:noncenteredpar}) is a standard Bayesian regression model for $\alphav$
  with a  conjugate  prior.

\section{Continuous shrinkage priors for sparse TVP models \label{ssm:sec:shrink}}

\subsection{From the ridge prior to continuous shrinkage priors \label{ssm:sec:priors}}

The ridge prior (\ref{ssm:normal}) for $ \sqrt \theta _j$ can be rewritten in the following way,
 \begin{eqnarray} \label{ssm:single}
  \sqrt \theta _j|\xiF_j 
  \sim \Normal{0,\comment{\tau} 
  \xiF_j} \quad \Leftrightarrow  \quad \theta_j |\xiF_j 
  \sim  \Gammad{\frac{1}{2}, \comment{\frac{1}{ 2 \comment{\tau} \xiF_j }}},
 \end{eqnarray}
where   $\xiF_j = 1$ is a  fixed scale parameter and $\comment{\tau}$ 
controls the global level of shrinkage \comment{of $\theta_j$, since
$\Ew{\theta_j| \tau} = \tau$}. 
%
In a sparse state space model, we expect that only a fraction of the coefficients are indeed dynamic, while \comment{the remaining coefficients} are (nearly) constant. This prior perception should be reflected in the choice of the prior distribution 
of the unknown variances $\theta_1, \ldots, \theta_\betad$.
In this section, we discuss how to incorporate this information through continuous shrinkage priors.
 In Section~\ref{ssm:sec:BVS}, we discuss mixture priors, also called spike-and-slab priors, in the context of \comment{variance} selection.

Under the ridge prior \comment{(\ref{ssm:single})},
$\xiF_j \sim \dirac{1}$  \comment{follows} a  point mass prior on 1, which does not allow for any local adaptation.
Continuous shrinkage priors take the form of global-local shrinkage priors in the sense of \cite{pol-sco:shr},
where $\xiF_j$ follows a prior $p(\xiF_j)$ that encourages many small values, representing coefficients that are nearly constant, while at the same time some of the $\xiF_j$'s are allowed to take on larger values to represent coefficients that are indeed time-varying.

For univariate sparse state space and TVP models, \cite{bel-etal:hie} introduced
  the Bayesian Lasso prior \cite{par-cas:bay}, where $\xiF _j$ follows an exponential distribution:
\begin{align}   \label{ssm:lasso}
\sqrt \theta_j   | \comment{\xiF_j } 
\sim    \Normal{0, \comment{\tau} 
\xiF_j},  \qquad \xiF_j \sim \Exp{1}.
\end{align}
This  prior is extended by \cite{bit-fru:ach} to the normal\Pcm{-}gamma prior \cite{gri-bro:inf}, where the exponential
prior for $p(\xiF_j)$   is generalized to a gamma prior:
\begin{align}   \label{ssm:double}
\sqrt \theta_j   | \xiF _j 
\sim    \Normal{0,\comment{\tau} 
\xiF_j}, \qquad
\xiF_j |a^\xi \sim \Gammad{a^\xi, a^\xi }.
\end{align}
For both priors, $\comment{\tau}$ 
acts as a global shrinkage parameter in a similar manner as for the ridge
prior (\ref{ssm:single}), however
each innovation variance $\theta_j$ is mixed over its {\em own} (local) scale parameter $\xiF _j$, each of which follows an independent exponential (\ref{ssm:lasso}) or a gamma distribution (\ref{ssm:double}). Hence, the $\xiF_j$'s play the role of local (component specific) shrinkage parameters. 
(\ref{ssm:double}) obviously reduces to the Bayesian  Lasso prior for $a^\xi=1$, but encourages more prior shrinkage toward small values and, at the same time, more extreme values than the Bayesian Lasso prior for  $a^\xi< 1$.

The normal\Pcm{-}gamma prior \comment{(\ref{ssm:double}) for $\sqrt \theta_j $}   can be represented in
the following way as a \lq\lq double gamma\rq\rq\  on $\theta_j$  \cite{bit-fru:ach}:
\begin{align}   \label{ssm:doublegam}
\theta_j  | \xi_j^2\sim \Gammad{\frac{1}{2}, \frac{1}{2\xi_j^2}}, \quad
\xi_j^2| a^\xi 
\sim \Gammad{a^\xi, \frac{a^\xi \kappa_B^2}{2}},
\end{align}
where $\kappa^2_B = \comment{2/\tau}$.
\cite{cad-etal:tri} proposed an extension of the double gamma prior (\ref{ssm:doublegam}) to a triple gamma prior, where
another layer is added to the hierarchy:
\begin{align}   \label{ssm:TRiple}
\theta_j  | \xi_j^2\sim \Gammad{\frac{1}{2}, \frac{1}{2\xi_j^2}}, \quad
\xi_j^2| a^\xi, \kappa_j^ 2 \sim \Gammad{a^\xi, \frac{a^\xi \kappa_j^2}{2}}, \quad
\kappa_j^2| c^\xi 
\sim \Gammad{c^\xi, \frac{c^\xi}{\kappa^2_B}}.
\end{align}
The main difference to the double gamma prior is that the prior scale of the $\xi^2_j$'s is not identical,
as  each local parameter $\xi^2_j$ depends on yet another local scale parameter $\kappa_j^2$. \comment{A similar prior 
is  applied to the initial expectations $\beta_j$:}
 \begin{align} \label{ssm:tpbnbeta}
\comment{ \beta_j  | \lambda_j \sim \Normal{0,  \lambda_j},  \quad
 \lambda_j|\tau_j^2  \sim \Gammad{a^\tau, \tau_j^2}, \quad
\tau_j^2  \sim \Gammad{c^\tau,\frac{2 c^\tau}{a^\tau \lambda_B^2}}.}
\end{align}
\cite{cad-etal:tri} show
that the triple gamma prior (\ref{ssm:TRiple}) can be represented as  a global-local shrinkage prior in the sense of  \cite{pol-sco:loc}, \comment{with the local shrinkage parameter $\xiF _j$ arising from  an $\Fd{2a^\xi, 2 c^\xi}$ distribution:}
\begin{align}
		\label{ssm:repF} & \sqrt \theta_j  |  \xiF _j 
 \sim \Normal{0, \comment{\tau} \xiF _j}, 
 \quad   \xiF _j | a^\xi,  c^\xi   \sim \Fd{2a^\xi, 2 c^\xi},
	\end{align}
with global shrinkage parameter $\comment{\tau=2/\kappa_B^2}$.
An interesting special case of the triple gamma is the horseshoe prior \cite{car-etal:hor}
 which results for $a^\xi=c^\xi =1/2$,
since  $\xiF _j    \sim \Fd{1,1}$ implies that $ \xiFsqr _j  \sim \Studentnu{1}$.
\remove{\begin{eqnarray*} 
\sqrt \theta_j  |   \xiF _j  \sim \Normal{0, \tau^2  \xiF _j },  \quad    \xiFsqr _j  \sim \Studentnu{1}.
\end{eqnarray*}}
\cite{cad-etal:tri} show that many other well-known shrinkage priors introduced in a regression context \comment{are} 
special cases of the triple gamma\Pc{,} which itself can be regarded as an application of the normal-gamma-gamma prior  \cite{gri-bro:hie} to  variance selection in the non-centered parametrization (\ref{ssm:eq:noncenteredpar}).

\begin{figure}[t!]
	\centering
\includegraphics[width=0.9\textwidth]{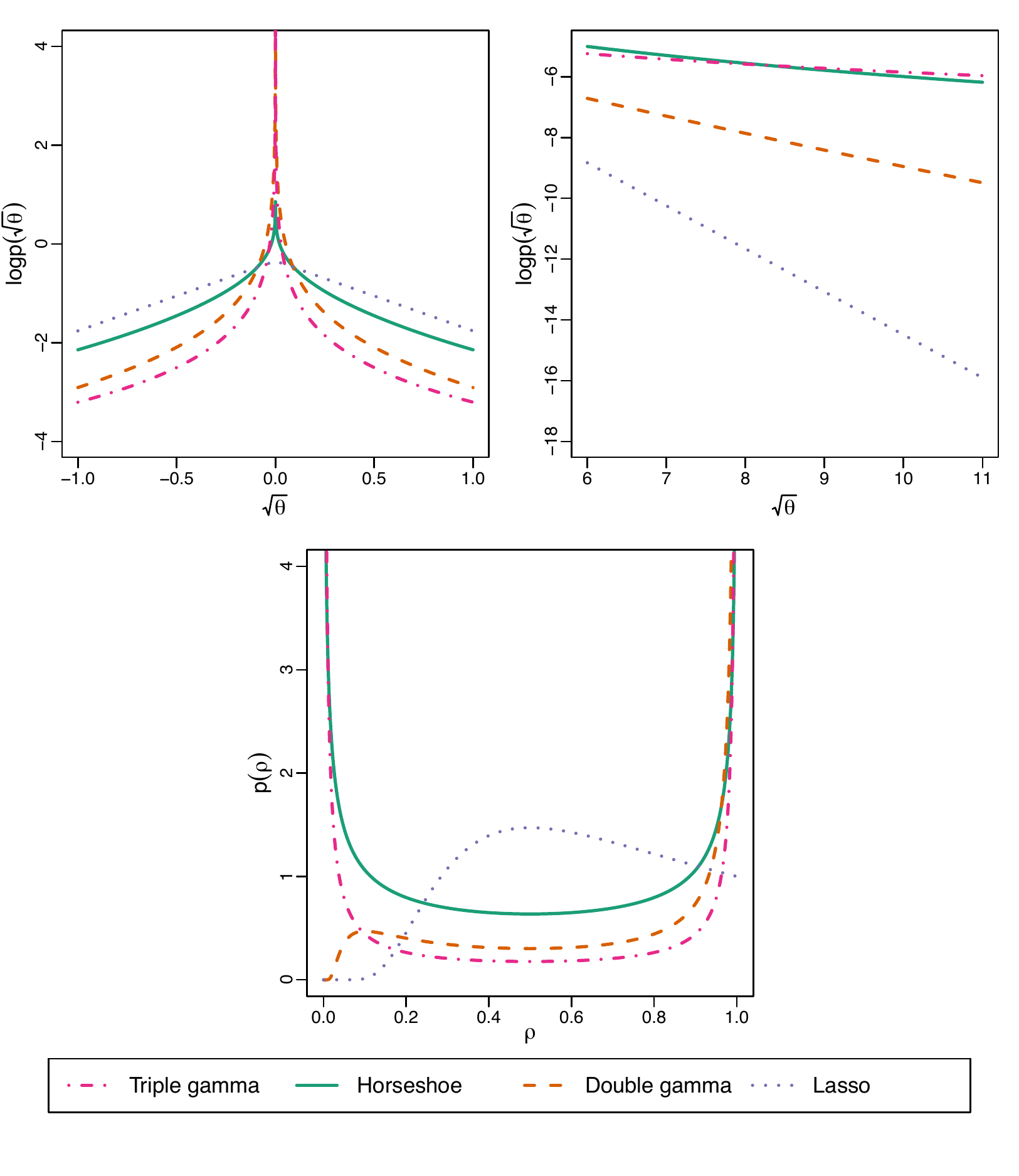}
	\caption{
Spike (top left-hand side) and  tail (top right-hand side)  of the marginal prior $p(\sqrt \theta_j)$
 and corresponding shrinkage profiles $p(\kappashr_j )$ (bottom) under the triple gamma prior with $a^\xi=c^\xi=0.1$  in comparison to the horseshoe prior, the double gamma prior with $a^\xi = 0.1$ and the Lasso prior. \comment{$\tau=1$ ($\kappa^2_B = 2$)} for all prior specifications.}
	\label{ssm:fig3}
\end{figure}


Among other representations,
  \comment{the triple gamma prior} has a representation as a generalized beta  mixture prior
introduced  by \cite{arm-etal:gen_bet}  for variable selection in regression models:
\begin{align} \label{ssm:densFrho}
\sqrt \theta_j   | \kappashr_j \sim \Normal{0, 1/\kappashr_j-1}, \quad
\kappashr_j| a^\xi, c^\xi, \phixi \sim \TPB{a^\xi, c^\xi, \phixi},
\end{align}
where  \comment{$\phixi=2 c^\xi /(\kappa_B^2 a^\xi)= \tau c^\xi /a^\xi $} and $\TPB{a^\xi, c^\xi, \phixi}$ is the three-parameter beta  distribution.
\remove{with density:
\begin{eqnarray*}
p(\kappashr_j )= \frac{1 }{\Betafun{a^\xi,c^\xi}}(\phixi) ^{c^\xi}  \kappashr_j   ^{c^\xi -1}
(1- \kappashr_j )  ^{a^\xi -1}   \left( 1+ (\phixi-1)\kappashr_j \right) ^{-(a^\xi+c^\xi)}.
\end{eqnarray*}}%
 \comment{This relationship
\Pc{makes it possible} to investigate the  shrinkage profile $p(\kappashr_j )$ of the triple gamma prior.
Figure~\ref{ssm:fig3} \Pc{contrasts} a triple gamma prior with   $a^\xi=c^\xi=0.1$  with
\Pc{a few} of its \Pc{special or limiting} cases, showing the behaviour around the origin, \Pc{in} the tails\Pc{,} as well as the shrinkage profile\Pc{s}.}

\comment{The \Pc{graphical representation} of the triple gamma prior in Figure~\ref{ssm:fig3} is based on
\cite{cad-etal:tri} who prove the following closed form expression for
the marginal prior $p(\sqrt \theta_j  |\phixi, a^\xi, c^\xi)$:}
		\begin{align}  \label{theo1pd}
		p(\sqrt \theta_j |\phixi, a^\xi, c^\xi) =   \frac{\Gamfun{c^\xi+{\frac{1}{2}}}}{\sqrt{2\pi \phixi} \Betafun{a^\xi,c^\xi}}
		\Uhyp{c^\xi + { \frac{1}{2}}, \frac{3}{2} - a^\xi, \frac{\theta_j}{2 \phixi} },
		\end{align}
		\comment{where $\Uhyp{a,b,z}=\int_0^\infty e^{-zt} t^{a-1}(1+t)^{b-a-1} dt$ is the confluent hyper-geometric function of the second kind.}

The parameter $a^\xi$ and $c^\xi$ control, respectively, the behaviour of this shrinkage prior at the origin and in the tails.
\cite{cad-etal:tri} prove that the triple gamma prior 
has \comment{an infinite spike} 
at the origin, if  $a^\xi \leq 0.5$, where for $a^\xi < 0.5$ and for small values of $\sqrt \theta_j $:
		\begin{align*}
		p(\sqrt \theta_j |\phixi, a^\xi, c^\xi) =   \frac{\Gamfun{\frac{1}{2}- a^\xi }}{\sqrt{\pi}  (2 \phixi) ^{a^\xi} \Betafun{a^\xi,c^\xi}} \left( \frac{1}{\sqrt \theta_j } \right)^{1 - 2 a^\xi} + O(1).
		\end{align*}
 Hence, the \comment{infinite spike} 
 is more pronounced, the closer $a^\xi $ \comment{is} to 0.
\comment{As  $\sqrt \theta_j  \rightarrow \infty$,} the triple gamma prior
has polynomial tails, with the shape parameter $c^\xi$ controlling the tail index:
		\begin{align*}
		p(\sqrt \theta_j  |\phixi, a^\xi, c^\xi) =
		\frac{\Gamfun{c^\xi+{\frac{1}{2}}} (2 \phixi) ^{c^\xi}}{\sqrt{\pi} \Betafun{a^\xi,c^\xi}}
		\left( \frac{1}{\sqrt \theta_j } \right)^{ 2 c^\xi+1} \left[1  + O \left( \frac{1}{\theta_j}\right) \right].
		\end{align*}

\paragraph*{Choosing the hyperparameters}

A challenging question is how to choose the parameters
	$a^\xi$,  $ c^\xi$  and\remove{ $\kappa_B^2$ or}
	$\phixi$  of  the triple gamma prior in the context of variance selection  for TVP models.
	\remove{In the same vein, the shrinkage parameters
	$a^\tau$, $ c^\tau$  and $\lambda_B^2$
 for the prior (\ref{ssm:tpbnbeta}) of the initial expectation $\beta_j$ have to be selected.}
	In high-dimensional settings it is appealing to have a prior that addresses two major issues:
	first, high concentration around the origin to favor strong shrinkage of small
	variances toward zero; second, heavy tails to introduce  robustness  to  large  variances and to avoid over-shrinkage.
	For the triple gamma prior, both issues are addressed
	through the choice of $a^\xi$ and   $ c^\xi$.

\comment{$a^\xi$ and $c^\xi$  can be fixed\Pcm{,} as for the Lasso 
and the horseshoe prior\Pcm{,}} 
or estimated from the data under a \comment{suitable} prior. 
\cite{cad-etal:tri}, e.g., assume that
\begin{align}   \label{ssm:prioraxi}
	2 a^\xi \sim \Betadis{\alpha_{a^\xi},  \beta_{a^\xi}} ,
	\qquad 2 c^\xi \sim  \Betadis{\alpha_{c^\xi},  \beta_{c^\xi}},
	\end{align}
restricting the support of $a^\xi $ and  $ c^\xi$ to $(0, 0.5)$, \comment{ensuring that the triple gamma prior is} more aggressive than the horseshoe prior.

	Ideally, one \comment{should} place a hyperprior distribution on the  global shrinkage parameter
 \comment{$\phixi $.} 
	Such a hierarchical triple gamma prior  introduces dependence among the local shrinkage parameters $\xi^2_1, \ldots, \xi^2_p$  in (\ref{ssm:TRiple})
	and, consequently, among $\theta_1, \ldots, \theta_\betad$ in the joint (marginal) prior  $p(\theta_1, \ldots, \theta_\betad)$.
	Introducing such dependence is desirable in that it allows \Pcm{the prior} to \comment{adapt}
 the degree of variance sparsity in a TVP model \comment{to the data at hand.}\remove{meaning that the amount a variance is shrunken toward zero depends on how close the other  variances are to zero.}
 \comment{For a  triple gamma prior  with arbitrary $a^\xi$ and (finite) $c^\xi$, \cite{cad-etal:tri} assume \Pcm{the} following prior on $\phixi $:
	\begin{align}  \label{ssm:hypfinal}
\phixi   |  a^\xi , c^\xi   \sim \Betapr{c^\xi,a^\xi }.
	\end{align}}\remove{which implies that
 \begin{align}
 	\left.  \frac{\kappa_B^2}{2}  \right|  a^\xi , c^\xi   \sim \Fd{2a^\xi ,2 c^\xi} .
 \end{align}}%
\comment{\noindent Prior~(\ref{ssm:hypfinal})  reduces to $\phixi   |  a^\xi \sim  \Fd{2a^\xi ,2 a^\xi}$ for
$a^\xi = c^\xi $. Hence,
	for the  horseshoe prior, 
$\phixi \sim   \Fd{1,1} $
and the  global shrinkage parameter $\tau = \sqrt{\phixi}$   follows a Cauchy prior
as in   \cite{bha-etal:las,car-etal:han}.}
 	As shown by \cite{cad-etal:tri}, under this hyperprior,  the  triple gamma prior  exhibits behaviour similar to Bayesian Model Averaging (BMA)\Pc{,} with a uniform prior on an  appropriately defined model size, see Section~\ref{ssm:sec:compare}.

\comment{For infinite $c^\xi$,
hierarchical versions of  the Lasso   and the double gamma prior  in TVP models 	are
	based on a gamma prior for the global shrinkage parameter
$ \kappa_B^2$, $ \kappa_B^2  \sim \Gammad{d_1, d_2}$} \cite{bel-etal:hie,bit-fru:ach}. This leads to a heavy-tailed extension  of both priors,
	where each marginal density  $p(\sqrt \theta_j  |d_1, d_2)$  follows a triple gamma prior with the same parameter  $a^\xi$ (being equal to one for the Bayesian Lasso) and  tail index $c^\xi =d_1$. \comment{In this light,
	very small values of $d_1$ had to be applied in these papers  to ensure heavy tails
	of $p(\sqrt \theta_j  |d_1, d_2)$.}

%

\subsection{Efficient MCMC inference\label{ssm:sec:MCMC}}

The two-block Gibbs sampler outlined in  Section~\ref{ssm:sec:motivate} can be extended to perform
MCMC inference for continuous shrinkage priors by exploiting the normal scale mixture representation underlying any global-local shrinkage prior.

Assume, for illustration, that we want to apply a normal\Pcm{-}gamma prior for the initial expectations $\beta_j$ and a double gamma prior for $\theta_j$:
\begin{equation} \label{ssm:exam}
\begin{aligned}
& \beta_j  |  \comment{\lambda_j} \sim \Normal{0, \comment{\lambda_j}}, \quad &\comment{\lambda_j| a^\tau} \typo{,\lambda_B^2}
 \sim \Gammad{a^\tau, \frac{ a^\tau \lambda_B^2}{2}},\\
& \theta_j  | \xi_j^2 \sim \Gammad{\frac{1}{2}, \frac{1}{2\xi_j^2}}, \quad
& \xi_j^2| a^\xi \typo{, \kappa_B^ 2}
\sim \Gammad{a^\xi, \frac{a^\xi \kappa_B^2}{2}},
\end{aligned}
\end{equation}
with fixed global shrinkage parameters  $a^\tau$, $\lambda_B^2$, $a^\xi$ and $\kappa_B^ 2$.
In this case, we can run a three-block Gibbs sampler to draw (a) the latent state process  from
$p(\kfzm| \typo{\alphav}, 
\sigma^2,\ym)$, (b) the model parameter
 \PcS{$\alphav=(\beta_1, \ldots, \beta_\betad ,\sqrt  \theta_1, \ldots, \sqrt \theta_\betad)$} from $p(\typo{\alphav}, 
 \sigma^2| \comment{\lambdav}, \xiv, \kfzm,\ym)$ conditional on knowing the local scale parameters $\comment{\lambdav=(\lambda_1, \ldots, \lambda_\betad)} $ and $\xiv=(\xi_1^2, \ldots, \xi^2_\betad)$, and (c)
  the local scale parameters from $p( \comment{\lambda_j}| \beta_j,\lambda_B^2)$ and $p(\xi^2_j|\theta_j, \kappa_B^ 2)$ for $j=1, \ldots,\betad$.


Let us  consider step~(c) in more {detail}, since sampling  the local shrinkage parameter  from $\comment{\xi^2_j}| \theta_j, \kappa_B^ 2$ (and similarly from $ \comment{\lambda_j}| \lambda_B^2,\beta_j$) is less standard.
 The double gamma prior  $\theta_j|\xi_j^2$ in  (\ref{ssm:exam})
leads to a  \comment{density}  for $\xi^2_j$ given $\theta_j$ which is the  kernel of an inverse gamma
density. In combination with the gamma prior for $\xi^2_j|a^\xi, \kappa_B^2$ also appearing in  (\ref{ssm:exam}),
this  leads to a posterior  distribution arising from a
 generalized inverse Gaussian (GIG) distribution: $ \xi^2_j| \theta_j, a^\xi, \comment{\kappa_B^2}  \sim  \GIG{a^\xi-1/2}{a^\xi \comment{\kappa_B^2}}{\theta_j}$.
\remove{The generalized inverse Gaussian distribution, $Y \sim \GIG{p}{a}{b}, a > 0, b > 0$  is a three parameter family, with support on $y \in \mathbb{R}^+$, and a density given by
\begin{eqnarray*}
 f(y) = \frac{(a/b)^{p/2}}{2 K_p(\sqrt{ab})} y^{p-1} e^{-(a/2)y} e^{-b/(2y)},
 \end{eqnarray*}
where  $K_p(\cdot)$ is the modified Bessel function of the second kind.}
 A very stable generator from the GIG distribution is implemented in the  R-package {\tt GIGrvg} \cite{hoe-ley:gig}.

\paragraph{To center or to non-center?}

In step (a) and (b) of the three--block sampler \comment{described above}, we have the option to either work with the centered parametrization (\ref{ssm:eq:centeredpar}) or the non-centered  parametrization (\ref{ssm:eq:noncenteredpar}).
Regardless of the parametrization, sampling the state process is straightforward, using either
 FFBS \cite{car-koh:ong,fru:dat} or a one-block sampler such as ``all without a loop” (AWOL) \cite{bit-fru:ach,kas-fru:anc}.

In the centered parametrization, the conditional posterior  $\theta_j| \beta_{j0}, \ldots, \beta_{jT},\beta_j$ is again
 a GIG distribution,
since the gamma prior for $\theta_j$ in (\ref{ssm:exam}) is  combined with the
\comment{density} $p( \beta_{j0}, \ldots, \beta_{jT}| \theta_j, \beta_j)$\Pc{,} which is  the  kernel of an inverse gamma density.
However, like many  MCMC schemes which alternate between sampling from the full conditionals of the latent states and the model parameters, the resulting \comment{sampler suffers from  slow convergence and poor mixing}
 if some of the  true process variances are small or even zero. \remove{This is illustrated in the left-hand side of Figure~\ref{ssm:fig5}.}

 \remove{\begin{figure}[t!]
\centering
\includegraphics[width=0.9\linewidth]{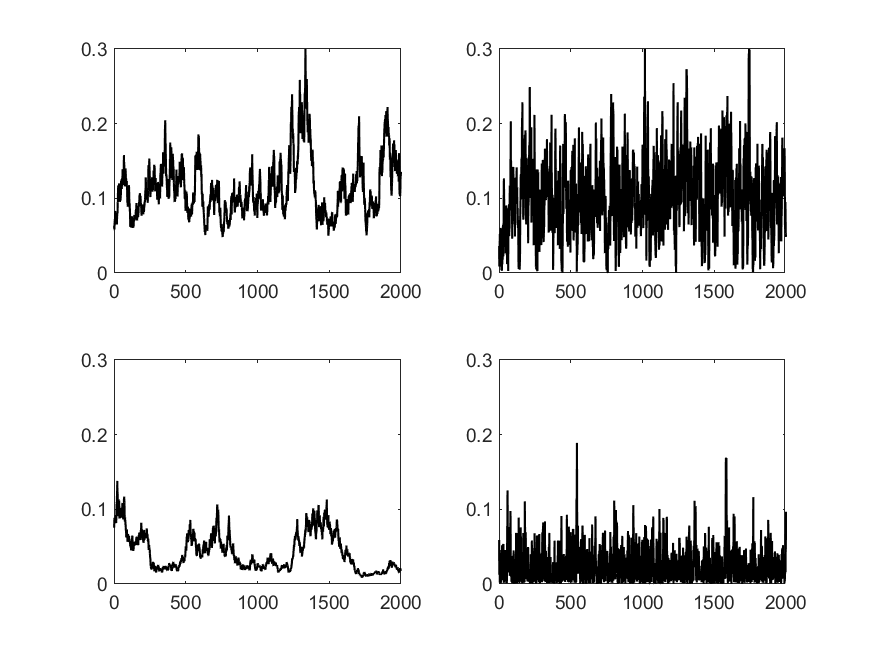}
\caption{\remove{The first 2000 MCMC draws (including burn-in) for $|\sqrt \theta_j|$ for the 
intercept  $\beta_{1t}$ (top)
and the 
coefficient $\beta_{2t}$ (bottom) for the simulated data
using the centered  (left-hand side) and  non-centered   (right-hand side) parametrization.}}
\label{ssm:fig5}
\end{figure}}

  As shown by \cite{fru-wag:sto}, MCMC estimation based on the non-centered parametrization   proves to be useful,  in particular if the process variances are close to \Pcm{zero}. \remove{, see the right-hand side of Figure~\ref{ssm:fig5}.}
Using the representation of the double gamma prior for $\theta_j$ as a \comment{conditionally normal prior,
$\sqrt \theta_j   | \xi_j^2  \sim    \Normal{0,\xi_j^2}$},
we obtain a joint Gaussian prior \comment{for} $\alphav=(\beta_1, \ldots, \beta_\betad ,\sqrt  \theta_1, \ldots, \sqrt \theta_\betad)$, where the
local shrinkage parameters $\comment{\lambdav}$ and $\xiv$ change the prior scale in a dynamic fashion during MCMC sampling.
Hence, in the non-centered  parametrization (\ref{ssm:eq:noncenteredpar}), conditional on
$\comment{\lambdav, \xiv}$ and the latent process
$\kfzm$\Pc{,}  
we are dealing with a Bayesian regression model under a non-conjugate analysis and
sampling from $p(\typo{\alphav}, 
\sigma^2| \comment{\lambdav}, \xiv, \kfzm,\ym)$ can be implemented as in Algorithm~\ref{ssm:algo1}.

  \cite{fru:com_eff}  discusses the relationship between the various
 parametrizations for  a simple TVP model  with $\betad=1$
 and the computational efficiency of the resulting MCMC samplers, see also  \cite{pap-etal:gen}.
 For TVP models with $\betad>1$,  MCMC estimation in the centered parametrization
is  preferable for all coefficients that are actually time-varying,
whereas the non-centered parametrization  is preferable for (nearly) constant coefficients.
   For practical time series analysis, both types of coefficients are likely to be present and choosing a computationally efficient parametrization in advance is not possible.  

  \remove{\begin{algorithm}[t!]
\remove{\begin{enumerate}
  \item[(a)] Sample the states $  \kfzm =( \tilde \betav_0, \ldots, \tilde \betav_T)$ in the non-centered parametrization   from
  the multivariate Gaussian posterior $ \kfzm| \betav,\kfQ, \sigma^2 $.
\item[(b)] Sample $\alphav  = (\beta_1, \dots,\beta_d, \sqrt \theta_1 ,\dots,\sqrt \theta_\betad )^{\top}$ from the multivariate Gaussian posterior
$p(\alphav| \kfzm, \tauv, \xiv, \sigma^2, \ym) $.
\item[(c)] ASIS: for each $j=1,\ldots,d$, redraw the constant coefficient $\beta_j$ and the square root of the process variance $\sqrt \theta_j $  through interweaving into the state equation of the centered parametrization:
    \begin{enumerate}
 \item[(c-1)]  Use the transformation (\ref{ssm:eq:solve}) to match  the latent process
 $\tilde \beta_{j0}, \ldots, \tilde \beta_{jT}$ into  the  latent process $\beta_{j0},\ldots, \beta_{jT}$ in the  centered parametrization.
\item[(c-2)] Update  $\beta_j$ and $\theta_j$ 
 in the centered parametrization by (a) sampling $\theta_j\new$ 
 from  the generalized inverse Gaussian  posterior $\theta_j|\beta_{j0},\ldots, \beta_{jT}, \beta_j, \xi^2_j$
  and (b) sampling
 $\beta_j \new$ from the Gaussian posterior  $  \beta_j | \beta_{j0}, \theta_j\new, \tau^2_j $.
   \item[(c-3)] Based on  $\sqrt \theta \new_j $ (using  the same  sign as the old value $\sqrt \theta_j$) and $\beta_j \new$, update the state process $\tilde \beta_{jt}$  in the non-centered parametrization  through the inverse of transformation (\ref{ssm:eq:solve}):
\begin{eqnarray*}
\tilde \beta _{jt} \new =  
(\beta_{jt} - \beta \new _j)/\sqrt \theta \new_j ,  \qquad  t=0,\ldots,T.
\end{eqnarray*}
\end{enumerate}
\item[(d)]  Sample the local shrinkage parameters  $ \tau_j |\beta_j, a^\tau, \lambda^2 $   and  $ \xi_j|  \theta_j, a^\xi, \kappa^2$, for $j=1,\ldots,d$,  from conditionally independent  generalized inverse Gaussian distributions.
\item[(e)] Sample the error variance $\sigma^2 |\kfzm, \alphav, \ym$ from an  inverse gamma distribution. \end{enumerate}}
 \caption{\remove{MCMC sampling including ASIS for a double gamma prior with fixed global shrinkage parameters  $a^\tau$, $\lambda_B^2$, $a^\xi$ and $\kappa_B^ 2$.}}\label{facsvalg}
\end{algorithm}}

  As shown by \cite{bit-fru:ach} in the context of TVP models,  these two data augmentation schemes can be combined   through the {\em ancillarity-sufficiency interweaving strategy} (ASIS)   introduced by  \cite{yu-men:cen} to obtain an efficient sampler  combining  the \lq\lq best of  both worlds\rq\rq .
ASIS provides a  principled way of  interweaving the centered and the non-centered parametrization of a TVP model
by  re-sampling certain parameters conditional on the latent variables in the alternative parametrization of the model.
More specifically,
  \cite{bit-fru:ach} 
   sample $\beta_j$ and $\sqrt \theta_j $ \comment{in the non-centered parametrization from
  the joint conditionally}  Gaussian distribution and interweave into the centered parametrization
  to resample $\theta_j$ from the conditional GIG distribution (and $\beta_j$ from yet another conditionally Gaussian distribution).
  This leads to an  MCMC sampling  scheme  which  increases posterior sampling efficiency considerably compared to sticking with  either of the two parametrizations throughout 
  sampling, while the additional computational cost of the interweaving step is minor.\remove{This is illustrated in Figure~\ref{ssm:fig5}. MCMC sampling based on ASIS is outlined  in Algorithm~\ref{facsvalg}.}
 \comment{ASIS was extended by \cite{cad-etal:tri} to the more general triple gamma prior.}

 \comment{MCMC sampling is extended by additional steps for hierarchical versions of the triple gamma prior, by sampling  all unknown global shrinkage parameters $a^\tau$, $c^\tau$, $\lambda_B^2$, $a^\xi$, $c^\xi$ and $\kappa_B^ 2$ from the appropriate conditional posterior distributions.} A full description of these algorithms can be found in  \cite{bit-fru:ach} for the double gamma prior and in \cite{cad-etal:tri} for the more general triple gamma prior.

\paragraph*{The \texttt{shrinkTVP} package}
The R package \texttt{shrinkTVP} \cite{kna-etal:shr_tim} offers efficient implementations of MCMC algorithms for TVP models with continuous shrinkage priors, specifically the triple gamma prior 
and its many special and limiting cases. It is designed to provide an easy entry point for fitting TVP models with shrinkage priors, while also giving more experienced users the option to adapt the model to their needs. The computationally demanding portions are written in C++ and then interfaced with R, combining the speed of compiled code with the ease-of-use of interpreted code.

\subsection{Application to US inflation modelling\label{ssm:sec:inf:shrink}}

In our application we model quarterly US inflation (1964:Q1 - 2015:Q4) as a generalized Philips curve with time-varying parameters in the spirit of \cite{koo-kor:for}. This means that inflation at time $t$ is modeled as
\begin{equation*} 
\begin{aligned}
& \kfx_{t}  = \kfx_{t-1} + \kfw_{t}, \qquad   \kfw_t  \sim \Normult{\betad}{\bfz, \kfQ},
\\
&y_t = \Xbeta_{t-1}\kfx_t  + \error_{t} , \qquad \error_{t} \sim \Normal{0,\comment{\sigma^2_t}},
\end{aligned}
\end{equation*}
where $y_t$ is inflation at time $t$, $\Xbeta_{t-1}$ is a set of \PcS{$p=18$} predictors including an intercept, exogenous variables from the previous time period and $y_{t-1}$ to \comment{$y_{t-\pVAR}$}, a series of lagged observations of inflation. For the application at hand we assume that \comment{$\pVAR = 3$}. The exogenous predictors included are broad and represent many different potential determinants of inflation. 
Table~\ref{ssm:table:data} offers an overview of the data, the sources used and the transformations applied to \comment{achieve (approximate) stationarity.} 
For this application we assume the error variance $\comment{\sigma^2_t}$
follows a stochastic volatility specification as in Section~\ref{ssm:sec:sv}.

\begin{table}[t!]
\centering
\caption{US inflation data description and sources}
\label{ssm:table:data}	
\small
\setlength{\tabcolsep}{3pt}
\begin{tabular}{@{}lllll@{}}
\toprule
Mnemonic & Description                                   & Database name & Source        & Tc    \\ \midrule
inf      & Consumer Price Index                          & CPI           & PHIL          & 4     \\
unemp    & Unemployment rate                             & RUC           & PHIL          & 1     \\
cons     & Real Personal Consumption Expenditures        & RCON          & PHIL          & 4     \\
dom\_inv & Real Gross Private Domestic Investment        & RINVRESID     & PHIL          & 4     \\
gdp      & Real GDP                                      & ROUTPUT       & PHIL          & 4     \\
hstarts  & Housing Starts                                & HSTARTS       & PHIL          & 3     \\
emp      & Nonfarm Payroll Employment                    & EMPLOY        & PHIL          & 4     \\
pmi      & ISM Manuf.: PMI Composite Index        		 & NAPM          & FRED          & 2     \\
treas    & 3m Treasury Bill: Secondary Market            & TB3MS         & FRED          & 1     \\
spread   & Spread 10-year T-Bond yield/3m T-Bill         & TB3MS - GS10  & FRED          & 1     \\
dow      & Dow Jones Industrial Average                  & UDJIAD1       & BCB           & 4     \\
m1       & M1 Money Stock                                & M118Q2        & PHIL          & 4     \\
exp      & Expected Changes in Inflation Rates           & -             & UoM           & 1     \\
napmpri  & NAPM Commodity Prices Index                   & NAPMPRI       & FRED          & 2     \\
napmsdi  & NAPM Vendor Deliveries Index                  & NAPMSDI       & FRED          & 2     \\
\bottomrule
\end{tabular}
\\
\vspace{2ex}
\raggedright{\footnotesize \textbf{Notes:} Tc refers to the transformation applied to the data. Let $\typo{z_{it}}$ be the original time series and $\typo{x_{it}}$ be the transformed time series, then 1 - no transformation, 2 - first difference, $\typo{x_{it}} = \typo{z_{it}} - \typo{z_{i,t-1}}$, 3 - logarithm, $\typo{x_{it}} = \log \typo{z_{it}} $, 4 - first difference of logarithm $\typo{x_{it}} = 100(\log \typo{z_{it}} -  \log\typo{z_{i,t-1}})$.
Sources are the Federal Reserve Bank of Philadelphia (PHIL),
the Federal Reserve Bank of St. Louis (FRED), the University of Michigan (UoM) and the Banco Central do Brasil (BCB).}
\end{table}

\begin{figure}[t!]
	\centering
	\includegraphics[width=1\textwidth]{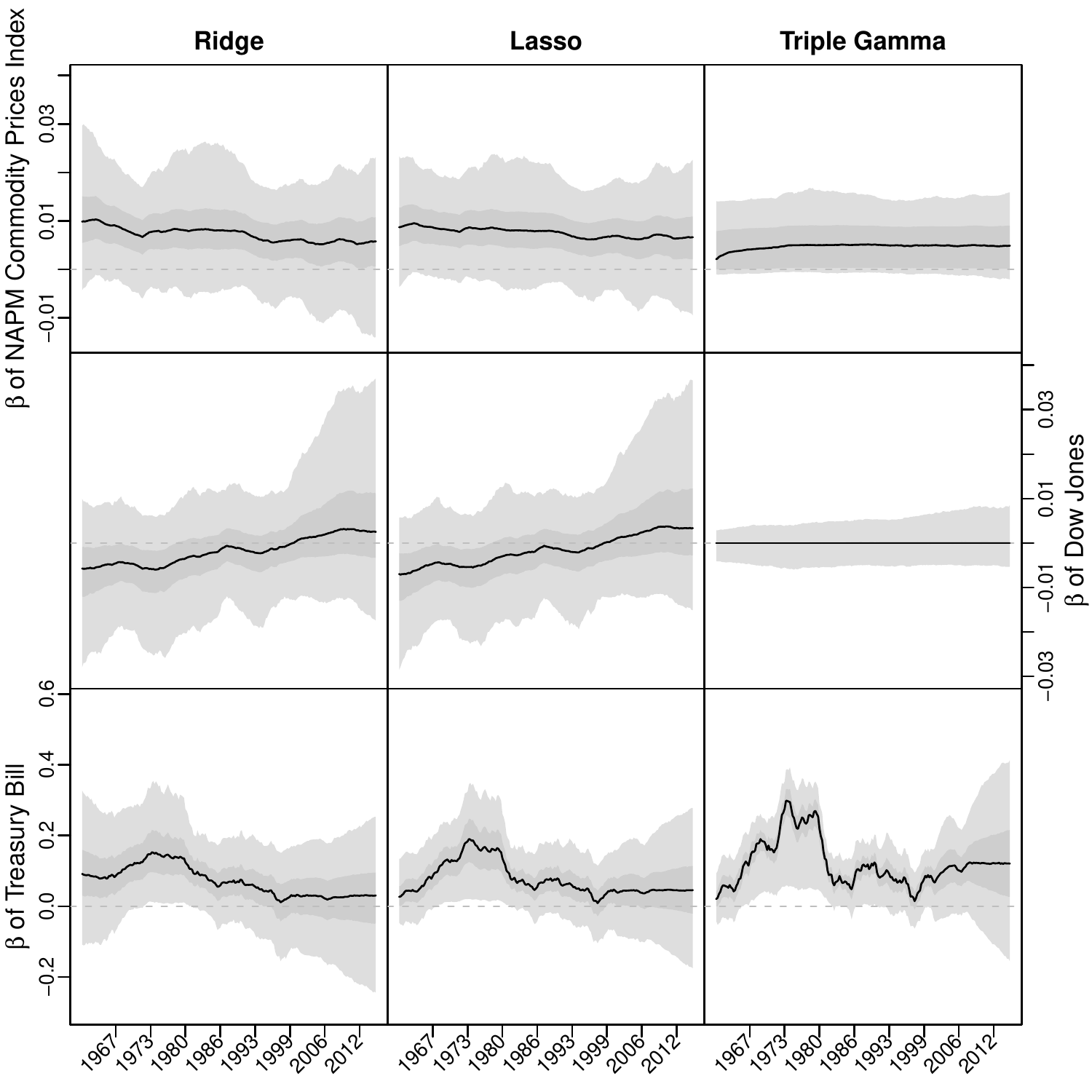}
	\caption{\comment{Recovery of} the time-varying parameters for the inflation data under the ridge prior, Lasso prior and triple gamma prior. The gray \comment{shaded regions} represent pointwise 95\% and 50\% credible intervals, respectively, while the black line represents the pointwise median.}
	\label{ssm:fig6}
\end{figure}

Three different priors are  placed on the expected initial values $\comment{\beta_1}, \ldots, \beta_p$ and on the variances of the innovations $\theta_1, \dots, \theta_p$, namely the ridge prior, as defined in equation~\eqref{ssm:normal}, the Lasso prior, as defined in equation~\eqref{ssm:lasso}, and the triple gamma prior, as defined in equation~\eqref{ssm:repF}. In the case of the Lasso prior, the global shrinkage parameters $\lambda_B^2$ and $\kappa_B^2$ are learned from the data under a gamma prior, specifically $\lambda_B^2 \sim \Gammad{0.001, 0.001}$ and $\kappa_B^2 \sim \Gammad{0.001, 0.001}$. In the triple gamma case, the hyperparameters are also learned from the data, under the priors defined in equations~\eqref{ssm:prioraxi} and \eqref{ssm:hypfinal}, with hyperparameter values $\alpha_{a^\xi} = \alpha_{a^\tau} = 5$ and $\beta_{a^\xi} = \beta_{a^\tau} = 10$.

Figure~\ref{ssm:fig6} shows how the three prior setups recovered the same states that were \comment{already} presented in \comment{Figure~\ref{ssm:fig2}}. 
\comment{While all three are noticeably smaller in scale than the states recovered under the inverse gamma prior,} they still differ in this regard as a consequence of the degree of shrinkage imposed, with the triple gamma prior imposing the most, followed by the Lasso prior and the ridge prior, in that order. This can be seen in the \comment{parameter for the Dow Jones} - the median is virtually zero under the triple gamma prior while displaying much more movement under the other two priors.
\comment{The parameter of the commodity prices index turns out to be significant, but practically constant under the triple Gamma prior, while the two other prior\Pc{s} \Pcm{also} assign considerable posterior mass to negative values.}
In the case of the parameter for the treasury bill, 
the most pronounced movement comes from the state estimated under the triple gamma prior, indicating that truly time-varying states are more likely to be picked up in such a sparse environment if the non time-varying parameters are effectively shrunken towards \comment{fixed ones}. 

\begin{figure}[t!]
	\centering
	\includegraphics[width=0.9\textwidth]{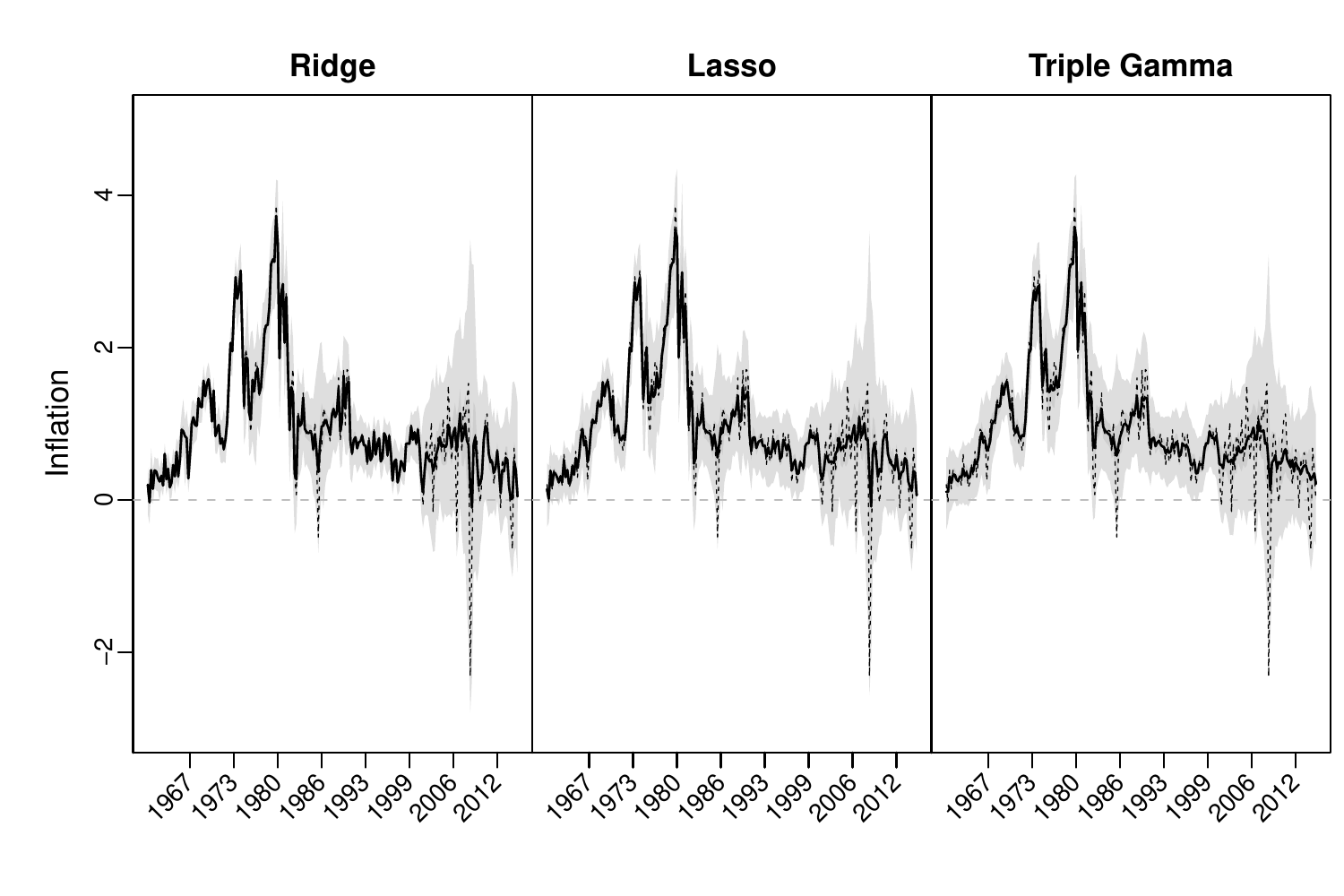} 
	\caption{Predicting the levels of inflation under the ridge prior, Lasso prior and triple gamma prior. The gray \comment{shaded regions} represent pointwise 95\% and 50\% credible intervals, respectively, while the solid black line represents the pointwise median. The dashed black line indicates the actual level of inflation.}
	\label{ssm:fig7}
\end{figure}

Another way to examine the effect that various levels of shrinkage have on the inference that follows is to look at the model implied predictions. Figure~\ref{ssm:fig7} plots the posterior predictive density of the three different models and contrasts these with the true levels of inflation. Two things are noteworthy: first, the stronger the shrinkage imposed by the prior, the less closely the median follows the true observation. This can be seen as shrinkage preventing the model from overfitting. Second, the error variance appears to be larger for the models with more shrinkage, as the spurious time variation in some parameters is dampened, leaving more of the variance to be soaked up by the error term. That this is beneficial for prediction can be seen in Section~\ref{ssm:sec:predict}.

\section{Spike-and-slab priors for sparse TVP models \label{ssm:sec:BVS}}

\subsection{From the ridge prior to spike-and-slab priors \label{ssm:sec:sps}}

A spike-and-slab prior  is a
  finite mixture distribution with two components, where one component (the {\em spike}) has much stronger global shrinkage than
   the second component (the {\em slab}). Such mixture shrinkage priors were introduced by
   \cite{geo-mcc:var,geo-mcc:app} for variable selection for regression models and aim \Pc{to identify} zero and non-zero regression effects. However,
  they are useful far beyond this problem and allow, \comment{for instance, parsimonious covariance
modelling for longitudinal data \cite{smi-koh:par},  covariance selection in random effects models \cite{fru-tue:bay} and robust random effects estimation \cite{fru-wag:bay}.}

\comment{Discrete spike-and-slab priors were introduced in 
 state space modeling} by \cite{fru-wag:sto}
 to achieve shrinkage of time-varying state variables toward fixed components.
In  \comment{TVP models},  such a prior is introduced  for the variance $\kfQc_j$ and reads
 \remove{\begin{eqnarray} \label{ssm:varrat}
    &&  \kfQc_j  \sim   (1-\pimix{\gamma}) p _{\spike} (\kfQc_j ) + \pimix{\gamma} p_{\slab} (\kfQc_j)  , \\
&& r_j =\frac{E_{\spike}(\kfQc_j)}{E_{\slab}(\kfQc_j)} << 1.    \nonumber
\end{eqnarray}
A special case of (\ref{ssm:varrat}) is the discrete spike-and-slab prior}  
$  \kfQc_j    \sim   (1-\pimix{\gamma})  \dirac{0} +  \pimix{\gamma}  p_{\slab} ( \kfQc_j)$,  
\comment{with the spike being a point measure at 0 and $p_{\slab} ( \kfQc_j)$ being the distribution in the slab.}
\cite{fru-wag:sto} introduced \Pc{the} following prior \comment{for the scale parameter  $\sqrt \theta_j $}
in the non-centered parametrization (\ref{ssm:eq:noncenteredpar}), with a ridge prior in the slab:
 \begin{eqnarray} \label{ssm:eq:DiscreteMix}
  \sqrt \theta_j | \sigma^2 \sim  (1-\pimix{\gamma})  \dirac{0} +  \pimix{\gamma}  \Normal{0,\sigma^2 B_{\gamma}}. \nonumber
\end{eqnarray}
\comment{With $\gamma_j$ being a binary indicator that separates the spike from the slab, $\pimix{\gamma}$ controls the prior occurrence of dynamic coefficients:}
\begin{eqnarray} \label{ssm:prgam}
\comment{\Prob{\gamma_j=1|\pimix{\gamma}}=\pimix{\gamma} .}
\end{eqnarray}
Again, this prior can be seen as an extension of the ridge prior, this time with
 a binary local scale parameter $\xiF_j=\gamma_j$  taking either the value 0 or 1:
 $ \comment{\sqrt \theta_j |\xiF_j=\gamma_j} 
  \sim   \Normal{0, \sigma^2 B_{\gamma} \gamma_j}$.
 A discrete spike-and-slab prior is also  applied 
to the initial expectation $\beta_j$:
\begin{eqnarray*}
 \beta_j| \sigma^2 \sim  (1-\pimix{\deltac} ) \dirac{0}  +  \pimix{\deltac} \Normal{0, \sigma^2  B_{\deltac}},
\end{eqnarray*}
with a corresponding binary indicator $\delta_j$ to separate the spike from the slab.
The dependence of the prior scale on  the error variance $\sigma^2$ in both priors
$ p(\beta_j| \sigma^2) $ and $p(\sqrt \theta_j | \sigma^2)$
allows sampling the indicators $\gamma_j$ and $\delta_j$ without conditioning on any model parameters, see
Section~\ref{ssm:sec:modsp}.

For a TVP model, the initial \comment{expectation $\beta_j$} is not identified if the parameter is actually
   time-varying. Therefore  it is not possible to discriminate between  $\delta_j=0$  and $\delta_j=1$, if   $\gamma_j=1$.
For this reason,  the following \comment{conditional prior for $\delta_j$ given
$\gamma_j$ is assumed:}
\begin{eqnarray*}
\Prob{\delta_j=1|\gamma_j=0,\pimix{\deltac}}=\pimix{\deltac} , \quad
\Prob{\delta_j=1|\gamma_j=1}=1, \nonumber 
\end{eqnarray*}
which rules out the possibility that $\delta_j=0$, while $\gamma_j=1$.
\comment{Combining this conditional prior with  (\ref{ssm:prgam})
leads to a joint prior for
 each pair $(\delta_j,\gamma_j)$ which has three possible realizations:}
  \begin{equation} \label{priordg}
  \begin{aligned}
& \Prob{\delta_j=0,\gamma_j=0}
=  (1-\pimix{\delta}) (1-\pimix{\gamma}) ,\\
& \Prob{\delta_j=1,\gamma_j=0} 
=  \pimix{\deltac}  (1-\pimix{\gamma}) ,  \\
& \Prob{\delta_j=1,\gamma_j=1} 
=  \pimix{\gamma}.
\end{aligned}
\end{equation}
%
\comment{As opposed to continuous priors, discrete spike-and-slab priors allow
explicit classification of the variables in a TVP model, based on 
$\deltac_j$ and $\gamma_j$:}
\begin{itemize}
  \item[(1)] A {\em dynamic} coefficient results if $\deltac_j =\gamma_j=1$, which implies $\betaci{j}\neq 0  $ and $\kfQc_j\neq 0$, in which case  $ \typo{\betaci{jt}} \neq  \typo{\betaci{j,t-1}} $ for all $t =1,\ldots,T$ and the  coefficient is allowed to change at each time point.
  \item[(2)] A {\em \comment{fixed}, non-zero} coefficient results if $\gamma_j=0$ but $\deltac_j = 1$, which implies $\betaci{j} \neq 0  $ while $\kfQc_j=0 $, in which case $ \typo{\betaci{jt}} =   \betaci{j} $ for all $t =1,\ldots,T$ and the
     coefficient is significant, but \comment{fixed}.
  \item[(3)] A {\em zero} coefficient results if $\deltac_j =\gamma_j=0$, which implies $ \betaci{j}=0$ and $\kfQc_j=0 $, in which case $ \betaci{tj} =   0 $ for all $t=1,\ldots,T$ and the coefficient is insignificant.
\end{itemize}
The probabilities given in (\ref{priordg})  are the prior probabilities for classifying coefficients into these
three categories. Based on this prior, in a fully Bayesian inference, the joint posterior distribution of $p(\deltav, \gammav |\ym)$ of all indicators $\deltav=
(\deltac_1, \ldots, \deltac_\betad)$ and  $\gammav=
(\gamma_1, \ldots, \gamma_\betad)$ is derived and can be used for posterior classification, e.g.
by deriving the model most often \Pc{visited,} or the median probability model. 


\paragraph*{Choosing hyperparameters for \comment{discrete} spike-and-slab priors}

First, the prior \PcS{probabilities} $ \pimix{\gamma}$ and $ \pimix{\deltac}$  to observe
 a dynamic or a \comment{constant} parameter, respectively,  have to be chosen.
As for standard variable selection, the strategy to fix  $ \pimix{\gamma}$ and $\pimix{\deltac}$
 is very informative on the model sizes.
  The numbers $\pdim{d}$, $\pdim{f}$ and $\pdim{0}$ of dynamic, \comment{constant}  and zero coefficients, respectively, are given by
\begin{eqnarray*}
\pdim{d}=\sum_{j=1} ^\betad \gamma_j, \quad \pdim{f}=\sum_{j=1} ^\betad \deltac_j (1- \gamma_j), \quad
 \pdim{0}=\sum_{j=1} ^\betad (1-\deltac_j) (1- \gamma_j).
\end{eqnarray*}
 \comment{Hence, apriori,   $\pdim{d}|\pimix{\gamma} \sim \Bino{\betad,\pimix{\gamma}}$, $\pdim{f}|\pimix{\deltac}, \pdim{d}  \sim \Bino{\betad - \pdim{d},\pimix{\deltac}}$,
 while $\pdim{0}$  given $\pdim{d}$ and $\pdim{f}$ is deterministic, $ \pdim{0}=\betad - (\pdim{d}+ \pdim{f})$.}

Alternatively, a hyperprior can be assumed for both probabilities in order to learn the desired
degree of sparsity from the data. Such a hierarchial prior allows more adaptation to the required level of sparsity and assumes that the prior probabilities $\pimix{\delta}$ and $\pimix{\gamma}$ are unknown, each following a beta distribution:
 \begin{eqnarray} \label{ssm:priodg}
\pimix{\delta} \sim \Betadis{a_0^\delta,b_0^\delta }, \qquad \pimix{\gamma} \sim \Betadis{a_0^\gamma,b_0^\gamma}.
\end{eqnarray}
Choosing  $a_0^\delta=b_0^\delta=1$ and $a_0^\gamma=b_0^\gamma=1$
implies that the prior on $\pdim{d}$ is uniform on $\{0, \ldots, \betad \}$,
while $\pdim{f}| \pdim{d}$   is uniform on $\{0, \ldots, \betad - \pdim{d} \}$.

Second, the prior \comment{in the  slab}  has to be specified.
For a discrete spike-and-slab prior,  all $\theta_j$s with $\gamma_j=0$ and
 all $\beta_j$s with $\delta_j=0$ are switched off in the non-centered model (\ref{ssm:eq:noncenteredpar}).
Hence, a prior has to be chosen for  the parameter $\betavred$  collecting all remaining non-zero $\beta_j$s and  $\sqrt \theta_j $s.
 \comment{Under a Gaussian slab distribution, such a prior reads}
\begin{eqnarray}  \label{ssm:piorssvv}
\betavred | \verror  \sim
\Normult{k}{\bfz, \verror  \tau 
\identy{k}},
\end{eqnarray}
  where $k=\pdim{f}+\typo{2 \pdim{d}}$.
 However, as  for variable selection in regression models, the choice of $ \tau$ 
 is influential in a higher-dimensional setting.
 A certain robustness is achieved by choosing a hierarchial Student-$t$ slab, where
  \begin{equation*} 
  \begin{aligned}
& \betaci{j}|\comment{\delta_j=1} 
  \sim  \Normal{0,\sigma^2 \lambda^2 /\tau_j^2}, \qquad
&\tau_j^2 & \sim \Gammad{a^{\tau},a^{\tau}}, \\
& \sqrt \theta_j | \comment{\gamma_j=1} 
  \sim \Normal{0,\sigma^2 \kappa^2/\xi_j^2},  \qquad
&\xi_j^2 &\sim \Gammad{a^{\xi},a^{\xi}},
\end{aligned}
\end{equation*}
with hyperpriors  $ \lambda ^2  \sim \Gammad{a^\lambda, a^\lambda}$  and $ \kappa ^2  \sim \Gammad{a^\kappa, a^\kappa}$ with small degrees of freedom, e.g. $a^{\tau}=a^{\xi}=a^{\lambda}=a^\kappa=0.5$.

Alternatively, \cite{fru-wag:sto} consider a  fractional prior which is commonly used in model selection\Pc{,} as it adapts the prior scale automatically in a way that guarantees model consistency \cite{oha:fra}. 
 \comment{For TVP models, \cite{fru-wag:sto} defined  a fractional prior for 
 $\betavred$  conditional on the \comment{latent process $\kfzm$} as
 $ p(\betavred|b,  \cdot) 
\propto \displaystyle p(\ym|\betavred,\verror, \kfzm )^b $.
This prior can be interpreted as the posterior of a non-informative prior combined with a small fraction $b$ 
of the complete data likelihood \PcS{$p(\ym|\betavred,\verror, \kfzm )$}.}


\subsection{Model space MCMC\label{ssm:sec:modsp}}

  \begin{algorithm}[t!]
\begin{enumerate}
\item[(a)] Sample \comment{indicators $\gammav=(\gamma_1,\ldots, \gamma_p)$ and
$\deltav=(\delta_1,\ldots, \delta_p)$}
from $p(\deltav,\gammav| \kfzm, \ym)$ conditional on the latent variables $\kfzm=(\tilde \kfx_{0}, \ldots, \tilde \kfx_{T})$;
\item[(b)] sample \comment{the  model parameters $\betavred$ and $\sigma^2$} conditional on $\kfzm$ and  $(\deltav,\gammav)$: 
\begin{enumerate}
\item[(b-1)]  sample $\sigma^2 $ from the inverse gamma density
   $\sigma^2 | \deltav,\gammav,\kfzm, \ym $
   \item[(b-2)] sample $\betavred$  from the multivariate Gaussian $\betavred| \sigma^2 , \kfzm, \ym $.
\end{enumerate}
  \item[(c)] sample  $\kfzm$
conditional on $\thmod=(\beta_1, \ldots, \beta_\betad , \sqrt\theta_1, \ldots, \sqrt\theta_\betad, \sigma^2 )$ from $\kfzm|\thmod, \ym $, using FFBS or AWOL.
\end{enumerate}
 \caption{Model space MCMC under a discrete spike-and slab prior with a conjugate Gaussian  slab.}\label{ssm:algo4}
\end{algorithm}

MCMC inference under discrete spike-and-slab priors is challenging, since the sampler is operating in a
very high-dimensional model space. Each of the $\betad$ covariates defines three types of coefficients,
hence the sampler needs to navigate through $3^\betad$ possible  models. The various steps of
model space MCMC are summarized in Algorithm~\ref{ssm:algo4} for the conjugate \comment{slab distribution (\ref{ssm:piorssvv}).}

Naturally, the most challenging part is Step~(a).
If $\betad$ is not too large,  then Step~(a) can be implemented as a full enumeration  Gibbs step
  by computing  the marginal likelihood  $p( \ym| \deltav,\gammav ,\kfzm)$ for all $3^\betad$  possible combinations of indicators, as illustrated by \cite{fru-wag:sto} for unobserved component state space models.
 Note that\Pcm{,} conditional on the latent process $\kfzm$,  $p( \ym| \deltav,\gammav , \kfzm)$ is  the marginal likelihood  of
 a constrained version of regression model (\ref{ssm:eq:noncenteredpar}) under the conjugate prior (\ref{ssm:piorssvv}) and therefore
 has a simple closed form.
 To derive the 
  posterior $p(\deltav,\gammav| \kfzm, \ym) \propto p( \ym| \deltav,\gammav , \kfzm)
 p(\deltav,\gammav) $,
 these marginal likelihoods are combined with the prior $p(\deltav,\gammav)$ \Pcm{for all models,}
  which is \Pc{available in closed form} even under the hierarchical prior (\ref{ssm:priodg}).

 In cases where such a full enumeration  Gibbs step becomes unfeasible because  $\betad$ is simply too
 large, Step~(a) can be implemented  as
 a single move sampler:
    loop randomly over all pairs of indicators
    $(\delta_j,\gamma_j), \PcS{j=1, \ldots, p},$ and propose to move from the current model
    $s=(\delta_j,\gamma_j)$ to a new model  $s\new=(\delta_j \new,\gamma_j \new)$ with
    probability $q_{s \rightarrow s\new}$. Accept $(\deltav,\gammav)\new$  with probability $\min(1,\alpha)$ where
\begin{eqnarray*} 
\alpha =\frac{p( \ym| (\deltav,\gammav)\new ,\kfzm) p((\deltav,\gammav)\new)  }
{ p( \ym|\deltav,\gammav ,\kfzm) p(\deltav,\gammav)} \times \frac{ q_{s\new \rightarrow s} }{  q_{s \rightarrow s\new} }.   
\end{eqnarray*}
 The art here is to design sensible moves. One strategy is to move with equal probability to one of the two alternative categories. For instance, if  currently  $\delta_j=\gamma_j=1$ defines a dynamic \PcS{coeffcient}, 
 then propose, respectively, with probability 0.5 to either move to a fixed coeffcient, where  $\delta_j\new = 0$ (while $\gamma_j\new =\gamma_j=1$)
 or to a zero coeffcient, where  $\delta_j\new= \gamma_j\new = 0$.
\comment{In general, moves involving a change from a fixed 
to a dynamic coefficient 
are not easily accepted. Given that $\gamma_j=0$,  the current latent path $\kfzm_j =(\tilde \beta_{j0}, \ldots, \tilde\beta_{jT}) $
was sampled from the prior
 $p(\kfzm_j)$  
which can be very different from  the smoothed posterior  
$p(\kfzm _j| \gamma\new_j=1, \ym) $, in particular if $T$ is large.}

Having updated the vector of indicators $(\deltav,\gammav)$, a modified version of Algorithm~\ref{ssm:algo1} is applied in Step~(b) and (c) \PcS{of Algorithm~\ref{ssm:algo4}} to sample the unconstrained model parameters $\betavred ,\verror | \comment{\kfzm}, \ym$ and $\comment{\kfzm} | \thmod ,  \ym$ in the restricted version of the non-centered parametrization. In particular, the sampling order is interchanged to obtain a valid sampler, since 
$(\deltav,\gammav)$ are updated without conditioning on the  \PcS{parameter $\thmod=(\beta_1, \ldots, \beta_\betad , \sqrt\theta_1, \ldots, \sqrt\theta_\betad, \sigma^2 )$.}

\begin{figure}[t!]
	\centering
	\includegraphics[width=0.8\textwidth]{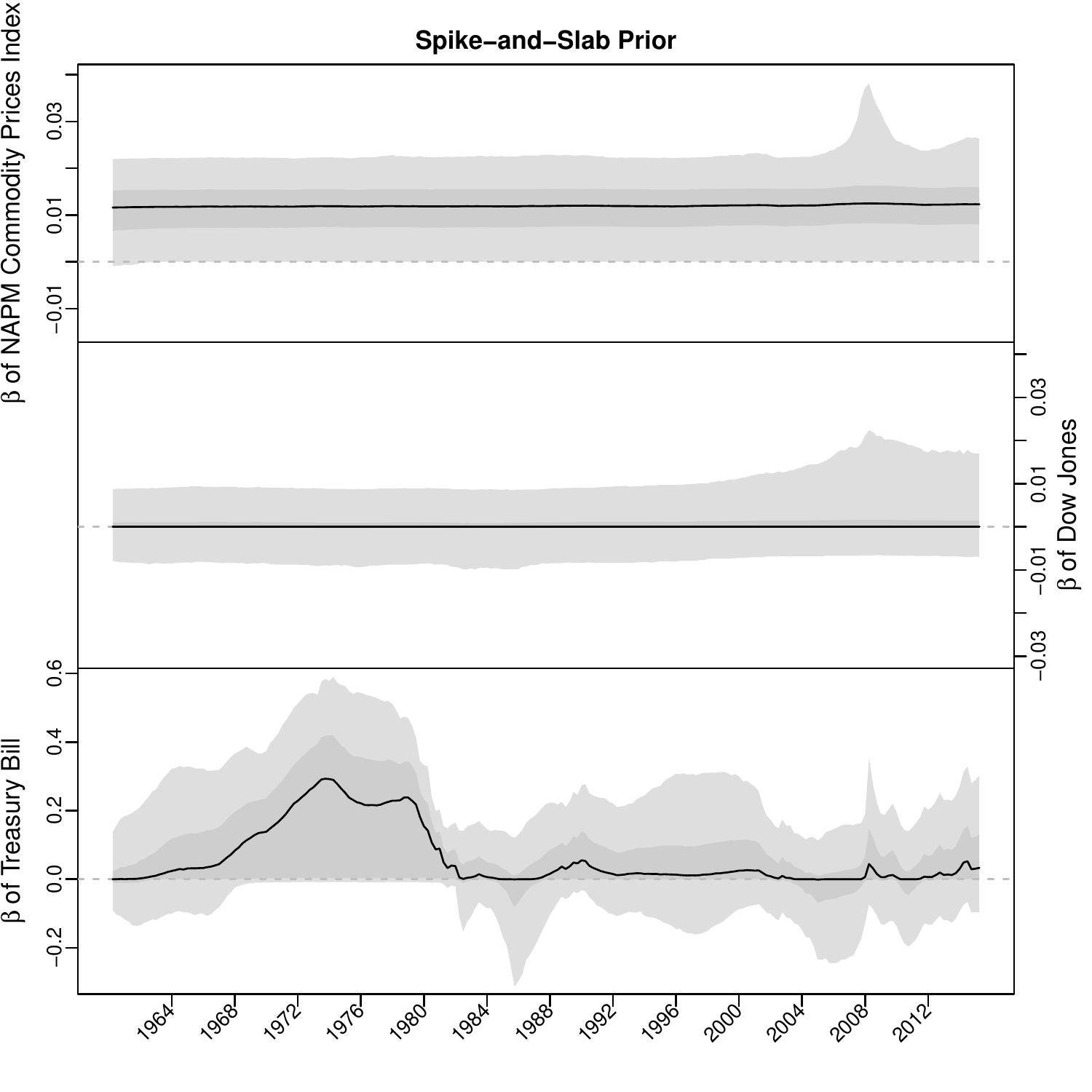} 
	\caption{\comment{Recovery of} the time-varying parameters for the inflation data under the discrete spike-hierarchical-Student-$t$-slab prior.
 The gray \comment{shaded regions} represent pointwise 95\% and 50\% credible intervals, respectively, while the black line represents the pointwise median.}
	\label{ssm:fig8}
\end{figure}

\subsection{Application to US inflation modelling\label{ssm:sec:inf:ssm}}

The analysis in Section~\ref{ssm:sec:inf:shrink} is extended,
 using discrete spike-and-slab priors for $\beta_j$ and $\sqrt \theta_j$  with following slab distributions:
  (1) Gaussian  with $\tau=1$, 
 (2) fractional priors with $b=10^{-4}$ and (3) hierarchial Student-$t$ 
 with  $a^{\tau}=a^{\xi}=a^{\lambda}=a^\kappa=0.5$.
 The hierarchical prior $\sigma^2|C_0 \sim \Gammainv{0.5,C_0}$, $C_0 \sim \Gammad{5,10/3}$
 is  assumed for the (homoscedastic) variance $\sigma^2$. The prior  of $\pimix{\delta}$ and $\pimix{\gamma}$ is chosen as in (\ref{ssm:priodg}) with $a_0^\delta=b_0^\delta=a_0^\gamma=1$ and  $b_0^\gamma=2$.

Model space MCMC sampling  was run for 100.000 iteration after a burn-in of 10.000.
\comment{Under the hierarchial Student-$t$ slab, the sampler exhibits
acceptance rates of around 20\% for all \Pc{classes} of moves\Pc{. This indicates} relatively good
performance,
given that the latent variables $\kfzm$ are unobserved and  imputed under the old indicators. }
 For Gaussian and fractional slabs, the average acceptance rate of moves between fixed and dynamic components was less than 5\%. To verify convergence, the sampler was run twice,
 starting either from a full TVP model with all $\gamma_j$s equal to 1 or from a standard regression model with all $\gamma_j$s equal to 0.
 Under the Student-$t$ slab, we found high concordance between the models sampled by both chains after burn-in.
 Under Gaussian and fractional slabs, however, the two chains were sampling totally different models, depending on the starting value.

\comment{The time-varying parameters  recovered under \Pc{the} hierarchical  Student-$t$ slab are shown in Figure~\ref{ssm:fig8}. We see
a similar discrimination between a dynamic path (treasury bills), a \comment{constant} path (commodity \PcS{prices} index) and a zero path (Dow Jones) as we saw in Figure~\ref{ssm:fig6} under the triple \Pc{g}amma prior.
A more formal discrimination based on the sampled indicators $\delta_j$ and $\gamma_j$ will be performed in Section~\ref{ssm:sec:compare}.}

\section{Extensions \label{ssm:sec:mult}}


\subsection{Including stochastic volatility \label{ssm:sec:sv}}

\comment{Assuming a homoscedastic  error variance $\sigma^2$ in the observation equation of
 the TVP model (\ref{ssm:eq:centeredpar})  may} create spurious time-variation in the coefficients, as discussed \comment{by \cite{sim:mac}}.
To be robust against conditional heteroscedasticity,
 $\sigma^2_t$ is \comment{often} assumed to \Pc{be} time-varying over $t=1,\dots,T$:
 \begin{equation*} 
\begin{aligned}
& \kfx_{t}  = \kfx_{t-1} + \kfw_{t}, \qquad   \kfw_t  \sim \Normult{\betad}{\bfz, \kfQ},\\
& y_{t}=   \Xbeta_t \kfx_{t}  +  \error_{t} , \qquad \error_{t} \sim \Normal{0,\sigma^2_t}.  
\end{aligned}
\end{equation*}
For TVP models, it is common to assume a  stochastic volatility (SV) specification \cite{jac-etal:bay_ana},
where the log volatility $h_t = \log \sigma^2_t $ follows an AR(1) process:
\begin{eqnarray} \label{ssm:svht}
h_t | h_{t-1}, \mu, \phi, \sigma_\eta^2 \sim \Normal{\mu+\phi(h_{t-1}-\mu),\sigma^2_\eta}.
\end{eqnarray}
The unknown model parameters $\mu$, $\phi$, and $\sigma_\eta^2$ in \eqref{ssm:svht} and the entire latent
volatility process $\{h_0, h_1,\ldots, h_T\}$ are added to the set of unknown variables.
MCMC estimation is easily extended using the very efficient sampler developed by \cite{kas-fru:anc} and implemented in the R-package {\tt stochvol} \PcS{\cite{hos-kas:mod}.}
\remove{Priors on  $\mu$, $\phi$ and $\sigma^2_\eta$  are chosen  as
$\mu  \sim \mathcal{N}( b_\mu, B_\mu )$, $(\phi +1 )/{2} \sim \mathcal{B}(\aphi, \bphi)$,
and $\sigma^2_\eta \sim \mathcal{G}(1/2, 1/2 \Bsv )$,
with hyperparameters $b_\mu = 0$ , $B_\mu = 100$, $\aphi = 5$, $\bphi = 1.5$, and $\Bsv = 1$ as default.}

\subsection{Sparse TVP models for multivariate time series\label{sec:mult_TVP_model}}


The TVP model (\ref{ssm:eq:centeredpar}) introduced in Section~\ref{ssm:sec:intro}
for univariate time series can be
 easily extended to TVP models for multivariate time series.
Consider, as illustration, the following  TVP model for a $\dimy$-dimensional  time series  $\ym_{t}$,
\begin{eqnarray} \label{ssm:eq:TVPmult}
\ym_{t} &=& \Bm_t \xm_{t} +\errorv_{t},\qquad \errorv_{t} \sim  \Normult{\dimy}{\bfz,\Sigmam_{t}} ,   
\end{eqnarray}
where $\xm_{t}$ is a  \emph{column}  vector of $\betad$ regressors,  and $\Bm_t $ is a  time-varying  $\dimmat{\dimy}{\betad}$  matrix with coefficient $\beta_{ij,t}$  in row $i$ and column $j$, potentially containing structural zeros or constant values. 

\paragraph*{Sparse  TVP  Cholesky SV models}

 One example is the  sparse TVP  Cholesky SV model  \cite{bit-fru:ach}, which reads for  $\dimy = 3$:
 \begin{equation} \label{ssm:cholsv}
 \begin{aligned}
y_{1t} &= \varepsilon_{1t},  					&\varepsilon_{1t} 	\sim \Normal{0, \e^{h_{1t}}}, \\
y_{2t} &= \beta_{21,t}y_{1t} +  \varepsilon_{2t}, &\varepsilon_{2t}	\sim \Normal{0, \e^{h_{2t}}}, \\
y_{3t} &= \beta_{31,t}y_{1t} + \beta_{32,t}y_{2t} +  \varepsilon_{3t},  	&\varepsilon_{3t}	\sim \Normal{0, \e^{h_{3t}}},
\end{aligned}
 \end{equation}
where the log volatilities $h_{it}$, $i=1, \ldots, \dimy,$  \comment{follow} \PcS{$q$} independent SV \PcS{processes  as defined} in (\ref{ssm:svht}), with row specific parameters   $\mu_i$, $\phi_i$, and  $\sigma^2_{\eta,i}$.
System (\ref{ssm:cholsv}) \comment{consists} of three independent univariate TVP models,
 \comment{where} no intercept is present. 
 \comment{In the first row,}  no regressors are present \comment{either} and only the 
 log  volatility $h_{1t}$ has to be estimated.
 In the $i$-th  equation,   $i-1$ regressors  are present and
   $i-1$ time-varying  regression coefficients  $\beta_{ij,t}$  as well as the time-varying volatility $h_{it}$ need  to be estimated.
%
System (\ref{ssm:cholsv}) can be written as
\begin{equation*} 
 \ym_t \sim  \Normult{\dimy}{ \Bm_t \xm_{t},\Dm_t},
\end{equation*}
where $\Bm_t$ is a   $\dimy \times \dimy$  matrix with \PcS{time-varying coefficients} 
$\beta_{ij,t}$, which are 0 for $j\geq i$.
$\Dm_t=\Diag{\e^{h_{1t}},\ldots,\e^{h_{\dimy  t}}}$
 is a diagonal matrix and the $\dimy$-dimensional  
 vector
$\xm_{t}=(y_{1t},\ldots,y_{\dimy t})^\top$ is \PcS{equal to} 
$\ym_t$.

 It is possible to show that this system is equivalent to the assumption of a dynamic covariance matrix,
$ \ym_t\sim \Normult{\dimy}{\bfz,\Sigmam_t}$,
where $\Sigmam_t = \Am_t \Dm_t \trans{\Am_t}$ and the dynamic Cholesky factor $\Am_t$ is lower triangular with ones
on the main diagonal and related to $\Bm_t$ through \comment{$\Am_t=(\identy{\dimy}- \Bm_t)^{-1}$}.

Both in (\ref{ssm:cholsv}) as well as in the more general system (\ref{ssm:eq:TVPmult}),
the  
unconstrained time-varying coefficients  $\beta_{ij,t}$ are assumed to follow  independent random walks as in the univariate case:
\begin{eqnarray} \label{eq:betaij}
\beta_{ij,t} = \beta_{ij,t-1} + \omega_{ij,t}, \quad \omega_{ij,t} \sim \Normal{0,\theta_{ij}},
\end{eqnarray}
 with initial values 
$\beta_{ij,0} \sim \Normal{\beta_{ij},\theta_{ij}} $.
Each of the  \comment{time-varying}  coefficients $\beta_{ij,t}$ is potentially constant, with the corresponding  process variance  $\theta_{ij}$ being 0.  A constant coefficient  $\beta_{ij,t} = \beta_{ij}$  is potentially insignificant, in which case   $\beta_{ij}=0$.
Hence, as for the univariate case, discrete spike-and-slab priors as introduced in Section~\ref{ssm:sec:BVS} or
  continuous shrinkage \Pcm{priors} as introduced in Section~\ref{ssm:sec:shrink}  \comment{are imposed  on   the fixed regression coefficients  $\beta_{ij}$\Pcm{,} as well as  the  process variances  $\theta_{ij}$}\Pcm{. This defines} a sparse multivariate TVP model for
identifying  which of these scenarios holds for each coefficient  $\beta_{ij,t}$.

It is advantageous to introduce  (hierarchical) 
shrinkage priors which are independent row-wise.
\comment{For instance, \cite{bit-fru:ach}}, introduce a hierarchical double gamma prior for 
 $\theta_{ij}$ and a hierarchical normal\Pcm{-}gamma prior for 
 $\beta_{ij}$ \Pc{for each row $i$ of the \comment{TVP} Cholesky SV model}.\remove{
\begin{eqnarray*} 
 \beta_{ij}|\tau^2_{ij} \sim  \Normal{0,\tau^2_{ij}}, \quad \tau_{ij}^2|a_i^\tau ,\lambda_i^2 \sim  \Gammad{a_i^\tau,a_i^\tau \lambda_i^2/2},
 \end{eqnarray*}
 with hyperpriors on the global shrinkage parameters $\lambda_i^2$, $a_i^\tau $, $\kappa^2_i$ and $a_i^\xi$, i.e. $\lambda_i^2 \sim \Gammad{e_1,e_2}$,  $a_i^\tau \sim \Exp {\ataupr}$, $\kappa^2_i \sim \Gammad{d_1,d_2}$, and  $a_i^\xi \sim \Exp {\axipr}$.
   By choosing  $a_i^\tau =a^\xi_i=1$,  a hierarchical Bayesian Lasso prior
   results.}
 Alternatively, independent discrete spike-and-slab priors with row-specific
 inclusion probabilities  
 can be specified.
Any of these choices  leads  to prior independence across the  $\dimy$ rows of the system (\ref{ssm:eq:TVPmult}) and
   both model space MCMC as well as boosted  MCMC  can be applied row-wise to perform posterior inference.

\paragraph*{\comment{Sparse} TVP-VAR-SV  models}

Another important example are time-varying parameter vector autoregressive models  of order $\pVAR$ with stochastic volatility (TVP-VAR-SV), where the $\dimy$-dimensional  time series  $\ym_{t}$  is assumed to follow
\begin{align} \label{TVPVAR}
\ym_t =  \cm_t + \Phim_{1,t} \ym _{t-1}  
+ \ldots  \Phim_{\pVAR,t} \ym _{t-\pVAR}  +  \errorv_t, \qquad \errorv_t \sim \Normult{\MVAR}{\bm 0, \Sigmam_t},
\end{align}
where $\cm_t$ is the $\MVAR$-dimensional time-varying intercept,  $\Phim_{j,t}$, for $j = 1, \ldots, \pVAR $ is a $\MVAR \times \MVAR$ matrix of time-varying coefficients,  and $\Sigmam_t$ is the time-varying variance covariance matrix of the error term.
Since the influential paper of \cite{pri:tim}, 
this model
has become a benchmark for analyzing relationships between macroeconomic variables that evolve over time, see \cite{cha-eis:bay,eis-etal:sto,fel-etal:sop,koo-kor:lar,nak:tim},
 among many others. 

Since all $\MVAR$ equations share the same predictor \commentS{$\Xbeta_t = ( 1, \ym_{t -1}^\top, \ldots,\ym_{t-\pVAR}^\top)^\top$}
(a vector of length $\betad=\MVAR \pVAR+1$),  the  TVP-VAR-SV model can be written in a  compact notation exactly as in (\ref{ssm:eq:TVPmult}) with  matrix
\begin{eqnarray*}
\Bm_t= \left(\cm_t \,\, \Phim_{1,t} \,\, \cdots \,\,  \Phim_{\pVAR,t} \right).
\end{eqnarray*}
All 
coefficients $\beta_{ij,t}$ in $\Bm_t$ follow independent random walks as in (\ref{eq:betaij}) with
initial expectation $\beta_{ij}$ and process variance $\theta_{ij}$. Due to the high dimensional nature of the \comment{time-varying matrix $\Bm_t$, shrinkage priors are instrumental for efficient inference\Pc{,} even for moderately sized systems.}
For instance,  \cite{cad-etal:tri}  introduce independent hierarchical triple gamma priors for  
$\beta_{ij}$  and    $\theta_{ij}$
  in each row $i=1, \ldots,\dimy $ of the TVP-VAR-SV model and demonstrate considerable efficiency gain compared to other shrinkage priors, such as the Lasso.


 Since $\Sigmam_{t}$ \Pc{is} typically a full covariance matrix, the rows of the system \PcS{\eqref{TVPVAR}} are not independent,
 as the various components in  $\errorv_{t}$ are correlated.
Following \cite{fru-tue:bay},  \cite{cad-etal:tri}   use the Cholesky decomposition $\Sigmam_t = 
\Am_t \Dm_t \Am_t ^\top $ to represent the TVP-VAR-SV model as a triangular system with independent errors $\etav_t \sim \Normult{\MVAR}{\bm 0, \Dm_t} $. $\Am_t$ is  lower triangular with ones on the main diagonal and the unconstrained elements
 $a_{ij, t}$ in the $i$-th row and $j$-th column of $\Am_t$ \Pc{again} follow random walks, with their own set of
 shrinkages priors on the corresponding variances and initial expectations.

 The TVP-VAR-SV model \Pc{then} has a representation   as a system of $\MVAR$ univariate TVP models, e.g. for $\MVAR=3$:
 \begin{equation*}   
 \begin{aligned}
&y_{1t} = \Xbeta_t \bm \beta_{t}^1 + \eta_{1t},  \quad &\eta_{1t} \sim \Normal{0, {\sigma^2_{1t}}} ,\\
&y_{2t} = {\Xbeta_t} \bm \beta_{t}^2 + a_{21,t} \eta_{1t} + \eta_{2t}
,  \quad &\eta_{2t} \sim \Normal{0, \sigma^2_{2t}} , \\
&y_{3t} = {\Xbeta_t} \bm \beta_{t}^3 + a_{31,t} \eta_{1t} +  a_{32,t}\eta_{2t} + \eta_{3t},  \quad & \eta_{3t} \sim \Normal{0, \sigma^2_{3t}},
\end{aligned}
\end{equation*}
where $\bm \beta_{t}^i$  is the $i$th row of $\Bm_t$.
For $i>1$, the $i$th equation 
is a univariate TVP  model \comment{with} the residuals
$\eta_{1t}, \ldots, \eta_{i-1,t} $ of the preceding $i - 1$ equations \comment{serving} as explanatory variables. Nevertheless, the time-varying parameters  $\bm \beta_{t}^i$ in each row  can be estimated equation by equation 
\cite{cad-etal:tri}.

\comment{It should be noted that both models might be sensitive to the ordering of the variables of the multivariate outcome $\ym_t$, see 
\cite{kil-lue:str} for a thorough discussion.}

%





\subsection{Non-Gaussian outcomes\label{ssm:nonGau}}

\comment{While the discussion of this chapter \Pcm{is centered around} Gaussian time series,
all methods can be extended to non-Gaussian time series, as demonstrated in
\cite{fru-wag:sto}\Pcm{,} who also considered 
time series of small counts based on the Poisson distribution. The main idea is to
augment auxiliary latent variables
$\omegav$ such that conditional on $\omegav$ a Gaussian TVP model results. Variable and variance \Pc{selection} is then performed conditional on $\omegav$, while an additional step in the MCMC scheme  imputes $\omegav$ given the remaining variables.}

 \comment{Examples include the representation of student-$t$ 
 errors as scale mixtures of Gaussians\remove{where $\sigma_t^2 \sim
  \Gammainv{\nu/2,\sigma^2 \nu/2}$ are iid stochastic volatilities. Another example
  are } and binary time series, where the representation $d_t=\indic{y_t>0}$ leads to the conditionally Gaussian state space model (\ref{ssm:eq:centeredpar}).
A similar strategy is pursued in \cite{fru-wag:bay,wag-dul:bay} for non-Gaussian
random effects models and in \cite{wag:bay_est} for dynamic survival models, see also
\cite{bha-etal:hor_reg_mac} for a recent review on regularisation in complex and deep models.}

\subsection{\comment{Log predictive scores} for comparing shrinkage priors \label{ssm:sec:predict}}

Log predictive density scores (\LPS ) are a widely used scoring rule 
 to compare models; see, e.g., \cite{gne-raf:str}. 
 As shown by \cite{bit-fru:ach},  log predictive density scores are also a useful means of evaluating and comparing different shrinkage priors for TVP models.
It is  common in this framework to use the first $t_0$ time series observations  $\ytr=(\ym_{1},\ldots, \ym_{t_0})$ as a \lq\lq training sample\rq\rq , while
evaluation is performed for the remaining observations $\ym_{t_0+1}, \ldots, \ym_T$. 

 For univariate time series $y_t$, 
 $\LPS$ is defined as:
\begin{eqnarray*} 
\LPS = \log  p(y_{t_0+1}, \ldots, y_T| \ytr ) =
 \sum_{t=t_0+1}^T  \comment{\LPSo{t}}, \quad  \LPSo{t}= \comment{\log}\,  p(y_{t}|  \ym^{t-1} ).
\end{eqnarray*}
\comment{For each point in time, 
$\LPSo{t}$ 
analyzes the performance separately for each $y_t$  and is obtained  by evaluating  the one-step ahead   predictive density $  p(y_{t}|  \ym^{t-1}) $
 given observations $ \ym^{t-1}=(y_1,\ldots, y_{t-1})$ up to $t-1$  at the {\em observed} value $y_t$.} \comment{$\LPS $ is  an aggregated measure of   performance  for the entire time series.}
As shown by  \cite{fru:bay} in the context of  selecting time-varying and fixed components for a basic structural
  state space model,  $\LPS $ 
  can be 
  interpreted as
  a log marginal likelihood based on the training sample prior $p(\thmod|  \ytr)$,  since
  \begin{eqnarray*}
  p(y_{t_0+1}, \ldots, y_T| \ytr) = \int p(y_{t_0+1}, \ldots, y_T | \ytr,\thmod) p(\thmod|  \ytr) d \, \thmod,
\end{eqnarray*}
 where $\thmod = (\beta_1, \ldots,  \commentS{\beta_p}$,  $\sqrt \theta_1 ,\ldots, \commentS{\sqrt \theta_p} , \sigma^2)$ summarises the unknown model parameters.  Hence, log predictive density scores provide a coherent foundation  for  comparing
 the predictive power of different types of shrinkage priors.

Determining $\LPSo{t}$ for each $t =t_0+1, \ldots, T$ can be challenging computationally.
In \cite{bit-fru:ach}, a Gaussian mixture approximation,  called the \emph{conditionally optimal Kalman mixture approximation}, is introduced to determine $p(y_{t}|\ym^{t-1})$ independently for \comment{each} $t$,
based on $M$ draws $\thmod^{(m)}, m=1, \ldots, M$ from the posterior  distribution $p(\thmod|\ym^{t-1})$.
\remove{
\begin{equation*}
 p(y_{t}|\ym^{t-1}) = \int p(y_{t}|\ym^t,\thmod) p(\thmod|\ym^{t-1}) d \thmod
 \approx \frac{1}{M} \sum^M_{m=1} \mathnormal{f}_N \left(y_{t};   \hat{y}_{t}^{(m)},  {S}^{(m)}_{t}  \right),
\end{equation*}
where $ \hat{y}_{t}^{(m)}$ and $  S^{(m)}_{t}  $  are obtained  from the prediction step of a Kalman filter for each
posterior draw $\thmod^{(m)}$.}


\comment{The whole concept can be extended to multivariate time series by defining} 
\begin{eqnarray*} 
\LPS = \log  p(\ym_{t_0+1}, \ldots, \ym_T| \ytr ) =
 \sum_{t=t_0+1}^T  \LPSo{t}, \quad
 \LPS^*_{t} =  \log \, p(\ym_{t}|\ym^{t-1}).
\end{eqnarray*}
\comment{In a triangular system such as the  TVP Cholesky SV  model 
and the TVP-VAR-SV  model 
discussed in Section~\ref{sec:mult_TVP_model},
errors are uncorrelated and we can exploit that}
\begin{eqnarray*}  
 \LPSo{t} =   \sum_{i=1}^\dimy  \LPSo{i t}, \quad
\LPSo{i t} = \log \, p(y_{it}|y_{1 t}, \ldots, y_{i-1,t}, \ym^{t-1}) .
\end{eqnarray*}
Since we condition on {\em observed} values $y_{1t}, \ldots, y_{i-1,t}$ in equation $i$,
 $\LPSo{it}$ can be determined independently for each $t$ \Pcm{and} for each equation $i$.
This allows \Pc{one} to  fully exploit the computational power of modern parallel computing facilities.

\paragraph*{Application to inflation modelling}

\begin{figure}[t!]
	\centering
	\includegraphics[width=1\textwidth]{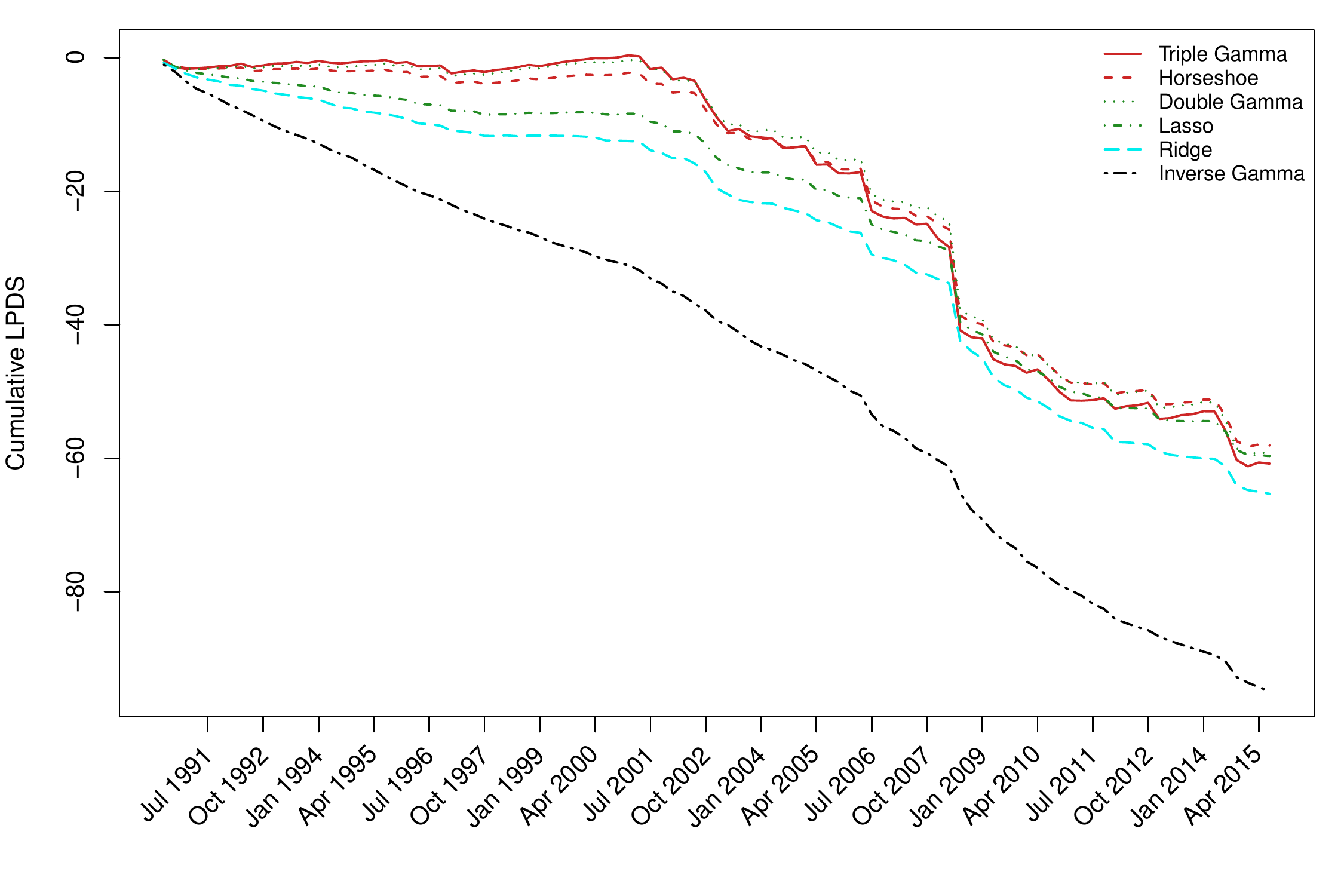}
	\caption{Cumulative LPDSs for the last 100 quarters of the inflation dataset introduced in Section~\ref{ssm:sec:inf:shrink}, for six different continuous (shrinkage) priors.}
	\label{ssm:fig9}
\end{figure}
To demonstrate the benefit that shrinkage provides with regards to out-of-sample prediction, we calculate one-step ahead LPDSs for the last 100 time points of the inflation dataset introduced in Section~\ref{ssm:sec:inf:shrink} and compute the cumulative sum.  \comment{Six} different prior choices are considered here: (1) the triple gamma prior, (2) the horseshoe prior, (3) the double gamma prior, (4) the Lasso prior, (5) the ridge prior and, finally, (6) the inverse gamma prior.
 Figure~\ref{ssm:fig9} displays the results, with higher numbers equating to better out-of-sample prediction. It is immediately obvious that the inverse gamma prior does not appear to be competitive in this regard. While it displays a high degree of in-sample fit (as evidenced by Figure~\ref{ssm:fig7}, Section~\ref{ssm:sec:inf:shrink}), the forecasting performance severely lags behind the other prior choices. Similarly, if not quite as drastically, the ridge prior does not forecast as well as the more strongly regularized approaches. The three priors with the most shrinkage, the triple gamma, the horseshoe and the double gamma, all perform comparably, while the Lasso prior initially lags behind, only to gain ground during the subprime mortgage crisis between 2007 and 2009.

\subsection{BMA versus continuous shrinkage priors\label{ssm:sec:compare}}

\begin{figure}[t!]
			\centering
			\includegraphics[width=1\textwidth]{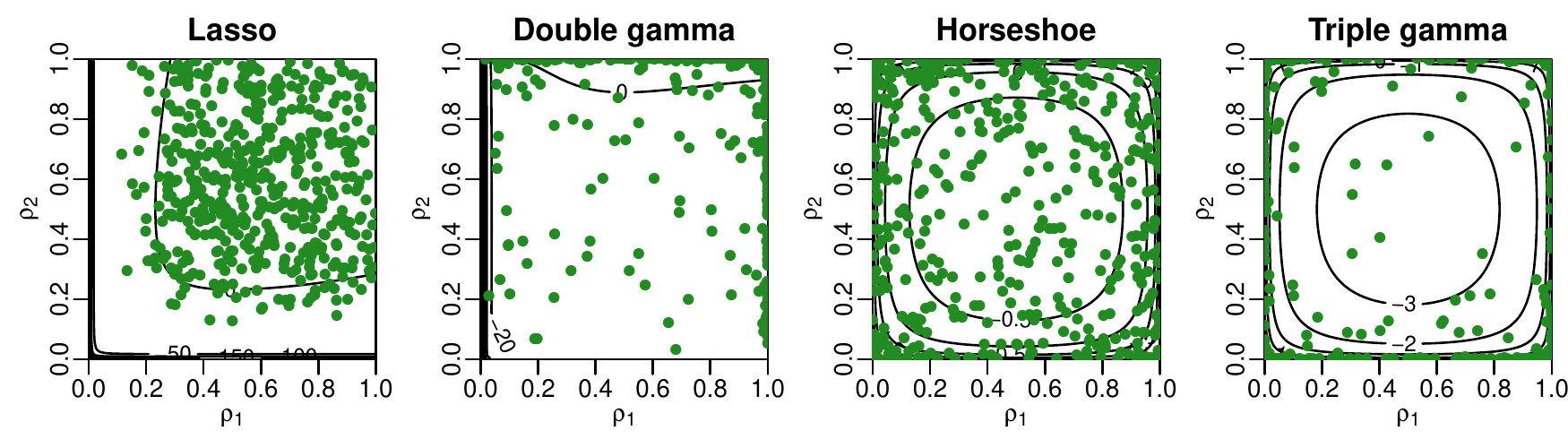}
			\caption{Bivariate shrinkage profile $p(\kappashr_1,\kappashr_2)$
				for (from left to right) the Lasso prior, the double gamma prior with $a^\xi = 0.1$, the horseshoe prior, and the triple gamma prior with $a^\xi=c^\xi=0.1$, with \PcS{$\tau=1$ ($\kappa^2_B = 2$)} for all the priors. The contour plots of the bivariate shrinkage profile are shown, together with 500 samples from the bivariate prior distribution of the shrinkage parameters.}
			\label{ssm:fig10}
		\end{figure}

	\comment{An interesting insight of \cite{cad-etal:tri}  is that the triple gamma prior  shows behaviour very similar to a discrete spike-and-slab prior as}
	  both $a^\xi$ and $c^\xi$ approach zero.
	This induces BMA-type behaviour on the joint shrinkage profile $p(\kappashr_1, \ldots, \kappashr_p)$,
	with \comment{an infinite} spike at all corner solutions, where some $\kappashr_j$ are very close to one, whereas
	the remaining ones are very close to zero.
 \comment{For illustration}, Figure~\ref{ssm:fig10} compares bivariate shrinkage profiles of various continuous shrinkage priors. The BMA-type behaviour of the triple gamma becomes evident through the large amount of mass placed in the four corners, with the overlayed \comment{500} samples from the prior following suit and clustering in those areas.

	Following   \cite{car-etal:han}, a natural way to perform variable selection in the continuous shrinkage prior framework is through thresholding.
	Specifically, when
	$ (1-\kappashr_j) > 0.5$, or $\kappashr_j< 0.5$, the variable is included, otherwise it is
not.\remove{Notice that this classification via thresholding makes perfect sense in the case of   a  triple gamma
	of which the horseshoe is a special case,
	but less so for  a Lasso or double gamma prior.}
	Notice that thresholding implies a prior on the model dimension
 $\pdim{d}$   defined as
	\begin{align*} 
	\pdim{d} = \sum_{j=1}^{\betad}\indicevent{ \kappashr_j< 0.5 } \sim
	\Bino{\betad,  \pimix{\gamma}}, \quad   \pimix{\gamma}= \Prob{\kappashr_j< 0.5 },
	\end{align*}
	where  $\kappashr_j 
\sim \TPB{a^\xi, c^\xi, \phi^\xi}$, see (\ref{ssm:densFrho}).
	The choice of the global shrinkage parameter  $\phi^\xi$ strongly impacts  the prior on  $\pdim{d}$.
	\comment{For a symmetric triple gamma prior with $a^\xi=c^\xi$ and $\phixi=1$ fixed}, for instance,\remove{that is $\kappa^2_B=2$,}  \comment{$\pimix{\gamma}=0.5$  and we obtain  $\pdim{d} \sim  \Bino{\betad, 0.5}$\Pcm{,}  regardless of $a^\xi$\Pcm{. This leads to }similar problems as with fixing  $\pimix{\gamma}=0.5$ for  \comment{a discrete spike-and-slab} prior.} 
 Placing a hyperprior on \comment{$\phi^\xi$}
  as discussed
in Section~\ref{ssm:sec:priors} is as vital for 
variance selection through
\comment{continuous shrinkage prior} as making $\pimix{\gamma}$  random is for \comment{a discrete spike-and-slab} prior. 
	\comment{\cite{car-etal:han} show that the hyperprior  for  $\phi^\xi$  defined  in (\ref{ssm:hypfinal})
leads \Pc{to} a uniform prior distribution on the model dimension $\pdim{d}$, since $\pimix{\gamma} \sim \Uniform{0,1}$ is uniformly distributed.}

\begin{table}
\caption{Classifying the coefficients for the inflation data in Table~\ref{ssm:table:data} into
 zero coefficients (z), constant coefficients (f) and time-varying coefficients (d)  under
 a discrete spike-and-slab  prior with hierarchical Student-$t$ slab. \label{ssm:table:class}}
\begin{center}
\begin{tabular}{lcccclccc}
\toprule
$\beta_{jt}$ & $\Prob{\mbox{\small z} |\ym }$ & $\Prob{\mbox{\small f} |\ym }$ & $\Prob{\mbox{\small d}|\ym }$
 && $\beta_{jt}$  & $\Prob{\mbox{\small z} |\ym }$ & $\Prob{\mbox{\small f} |\ym }$ & $\Prob{\mbox{\small d}|\ym }$\\[1pt]
 \hline
intercept &   0.30 &  0.41 &   0.29 && emp   &   0.38 &  0.42 &    0.20 \\
$y_{t-1}$ &   0.39 &  0.39 &   0.22 && pmi  &  0.57  &  0.35 &  0.08 \\
$y_{t-2}$ &   0.36 &  0.39 &   0.25 && treas &  0.03 &  0.11 &   {\bf 0.86} \\
$y_{t-3}$ &   0.28 &  0.35 &   0.37 && spread &   0.37 &  0.41 &   0.22 \\
unemp &   0.46 &  0.36 &   0.18 && dow  &   {\bf 0.61} &   0.30 &  0.09 \\
cons &   0.28 &  0.39 &   0.33 && m1 &   0.39 &  0.44 &   0.17 \\
dom\_inv &   0.58  &  0.34 &  0.08 && exp &   0.27 &  0.45 &   0.27 \\
gdp &   0.45 &  0.39 &   0.16 && napmpri &  0.05 &  {\bf 0.78} &   0.17 \\
hstarts   &   0.41 &  0.43 &   0.16 && napmsdi &   0.60 &  0.34 &  0.06\\
\bottomrule
\end{tabular}
\end{center}
\end{table}

\paragraph*{Application to US inflation modelling}

For illustration, we compare \comment{discrete} spike-and-slab priors and hierarchical
\comment{continuous} shrinkage priors with regard to classification of the time-varying parameters for the inflation data set introduced in Section~\ref{ssm:sec:inf:shrink}.  The posterior probabilities of each
coefficient to be either zero, fixed or dynamic  are estimated from the
 $M$ posterior draws of $(\delta_j \im{m},\gamma\im{m})$:
 \begin{eqnarray*}  
& \displaystyle\Prob{\beta_{jt} \mbox{ \small dynamic}|\ym }= \frac{1}{M} \sum_{m=1} ^M  \gamma_j\im{m},  \nonumber
 \,\, \Prob{\beta_{jt} \mbox{ \small fixed} |\ym }= \frac{1}{M} \sum_{m=1} ^M \deltac_j \im{m} (1- \gamma_j\im{m}) ,& 
\end{eqnarray*}
and $\Prob{\beta_{jt} \mbox{ \small zero} |\ym }= 1- \Prob{\beta_{jt} \mbox{ \small dynamic}|\ym }-
\Prob{\beta_{jt} \mbox{ \small fixed} |\ym }$.
The indicators $(\delta_j \im{m},\gamma\im{m})$ are an immediate outcome of the model space
MCMC sampler for the \comment{discrete} spike-and-slab prior and are derived for
\comment{continuous} shrinkage priors using 
thresholding as explained above.

According to this procedure, none of the coefficients is classified other than zero for the Lasso prior, which is not surprising in light of Figure~\ref{ssm:fig6}. Somewhat unexpectedly, the same classification results for the triple gamma prior, for which a clear visual distinction can be made in Figure~\ref{ssm:fig6} between the relatively  dynamic  coefficient of treasury bills and the \comment{other two} coefficients 
which are shrunken  toward \comment{a fixed coefficient}.

As opposed to this, the \comment{discrete} spike-and-slab prior shows more power to discriminate between the different types of coefficients for this specific data set. \comment{The corresponding classification probabilities
are reported for each coefficient in  Table~\ref{ssm:table:class} and  match the behaviour of the recovered time-varying \PcS{coefficients in Figure~\ref{ssm:fig8}.} 
More specifically, treasury bills is clearly classified as dynamic, the commodity \PcS{prices} index is classified as a having positive, but fixed effect on inflation, and the Dow Jones
is clearly classified as insignificant.}

\section{Discussion\label{section:ssm:conclude}}

\comment{This chapter illustrates the importance  of  variance selection for TVP models.
 If the true model underlying  a time series}  is sparse, with many coefficients being 
\comment{constant} or even zero, then a full-fledged TVP model might \Pc{quickly} \comment{overfit}.
 To avoid loss of statistical efficiency in parameter estimation and forecasting that goes hand-in-hand with the application of an overfitting model, we generally
 recommend to substitute  the popular inverse gamma  prior for  the process variances by suitable shrinkage priors.
 \comment{As  demonstrated in this chapter,}  shrinkage priors are indeed  able to automatically reduce time-varying  coefficients to \comment{constant or even insignificant} ones.

Within the class of continuous shrinkage priors, flexible priors such as hierarchical versions of the double gamma, the triple gamma or the horseshoe prior typically turn out to be preferable to less flexible priors such as the
hierarchical Lasso. These priors often show a comparable behaviour in terms of model comparison through log predictive density scores and they beat the inverse gamma prior by far. This was illustrated with an application to US inflation modelling  using a TVP Phillips curve.

Discrete spike-and-slab priors  are an attractive alternative to continuous shrinkage priors as they allow explicit classification of the time-varying coefficients into dynamic, \comment{constant and zero ones}. For continuous shrinkage priors, such a classification can be achieved only indirectly through thresholding \comment{and the appropriate choice of the truncation level is still an open issue for TVP models.} However, convergence problems with \comment{model space MCMC algorithms} are common with discrete spike-and-slab priors and the sampler might get stuck in different parts of the huge model space, depending on where \Pc{the algorithm is intialized}.
In our illustrative application, discrete spike-and-slab priors were more successful in classifying obviously time-varying coefficients than any continuous shrinkage prior, but only in combination with a 
Student-$t$  slab distribution. For other slab distributions, in particular Gaussian ones, severe convergence problems with trans-dimensional MCMC estimation were encountered.

A key limitation of any of the approaches
reviewed in this chapter is that they can only differentiate between parameters that are constantly time-varying or not time-varying at all. \comment{One could think of many scenarios}
in which a parameter may be required to be time-varying over a stretch of time and be constant elsewhere. The design of suitable {\em dynamic shrinkage priors} that are able to handle such a situation is cutting-edge research in the area of state space and TVP models. Very promising approaches toward dynamic shrinkage priors
were put forward by a number of author\Pc{s}, including
\comment{\cite{cas-etal:bay_non,kal-gri:tim,kow-etal:dyn,roc-mca:dyn}.}

A script to replicate select results from this chapter and instructions on how to download software routines in \texttt{R} is made available as part of the online supplement of this edited volume.








\printindex
\cleardoublepage
\end{document}